\documentclass{template_SCOR_article}

\def \ind{\mathds{1}}

\def \EE{\mathbb{E}}

\def \NN{\mathbb{N}}
\def \RR{\mathbb{R}}

\def \cD{\mathcal{D}}
\def \cE{\mathcal{E}}

\def \cI{\mathcal{I}}

\def \bx{\mathbf x}
\def \bX{\mathbf X}

\newcommand{\vero}[1]{{\color{black}#1}}
\newcommand{\verob}[1]{{\color{black}#1}}
\newcommand{\veroc}[1]{{\color{black}#1}}
\newcommand{\kev}[1]{{\color{black}#1}}

\title{Random Forest \veroc{based Qantile Oriented Sensitivity Analysis indices} estimation}
\date{\small \today}

\begin{document}

\maketitle

\begin{abstract}
\veroc{We propose a random forest based estimation procedure for Quantile Oriented Sensitivity Analysis - QOSA. In order to be efficient, a cross validation step on the leaf size of trees is required. Our full estimation procedure is tested on both simulated data and a real dataset.}
\end{abstract}

\veroc{\underline{Keywords:} 
Quantile Oriented Sensitivity Analysis, Random forest, Cross validation, Out of Bag samples.
}\ \\
\section{Introduction}\label{sec:1:introduction}

\veroc{Numerical models} are ubiquitous in various fields, such as aerospace, economy, environment or insurance, they allow to approximate the behavior of physical phenomenon. Their main advantage is that they replace expensive, or even unachievable, real-life experiments and thus provide knowledge about the natural system. The extremely faithful representation of reality, made possible thanks to the increase in computing power, also explains this widespread use. However, this accuracy is often synonymous of complexity, ultimately leading to a difficult interpretation \veroc{of models}. Besides, model inputs are usually uncertain due to a lack of information or the random nature of \veroc{ factors}, which means that the resulting output can be regarded as random. It is then important to assess the impact of this uncertainty on the model output. Global Sensitivity Analysis (GSA) methods solve these issues by \textit{studying how the uncertainty in the output of a model can be apportioned to different sources of uncertainty in the model inputs} \citep{saltelli2004sensitivity}. Hence, GSA allows to investigate input-ouput relationships by identifying the inputs that strongly influence the model response. Conversely, it may be of interest to see that although some inputs may not be very well established, they do not significantly contribute to output uncertainty.

Variance-based approaches are well-established and widely used for GSA. Among them, the sensitivity indices developed by \citet{sobol1993sensitivity} are very popular. This last method stands on the assumption that the inputs are independent. Under this hypothesis, the overall variance of a scalar output can be split down into different partial variances using the so-called \cite{Hoeffding48} decomposition. Then, the first-order Sobol' index quantifies the individual contribution of an input to the output variance while the total Sobol' index \citep{jansen1994monte,homma1996importance} measures the marginal and interaction effects. However, even if they are extremely popular and informative measures, variance-based approaches suffer from \veroc{some} limitation. Indeed, by definition, they study only the impact of the inputs on the expectation of the output \veroc{since they consider} the variance as distance measure. 

A new class of sensitivity indices, generalizing the first-order Sobol' index to other quantities of interest than the expectation, has been introduced in \citet{fort2016new}. These indices called Goal Oriented Sensitivity Analysis (GOSA) compare the minimum of a specific contrast function to its conditional counterpart when one of the inputs is fixed. The unconditional minimum being reached by the quantity of interest (for example a quantile). 

In this paper, we focus on Quantile Oriented Sensitivity Analysis (QOSA) measuring the impact of the inputs on the $\alpha$-quantile of the output distribution. \citet{browne2017estimate,maume2018estimation} introduced a statistical estimator of the first-order QOSA index based on a kernel approach. \veroc{\citet{kala2019quantile} defined} the second and higher order QOSA indices as well as a variance-like decomposition for quantiles in the case of independent inputs. \veroc{\citet{qosa-shapley} studied QOSA indices on various toy models in independent and dependent contexts. }\\
Despite these recents works, the question of the effective estimation of the first-order QOSA index remains open. Indeed, it turns out to be difficult to compute \veroc{them} in practice because  it requires an accurate estimate of either the conditional quantile of the output given an input, or the minimum of a conditional expectation of the output given an input. \citet{kala2019quantile} handles this feature with a brute force Monte-Carlo approach. As a matter of fact, for each value of an input, realizations of the other inputs are generated conditionally to the fixed value. Therefore, in this approach, the dependency structure of inputs has to be known, which is not always the case. Besides, the computational cost is too high to consider its use in an industrial context when dealing with costly models. \citet{browne2017estimate,maume2018estimation} developed kernel-based estimators to avoid this double-loop issue. But, when using a small dataset, their performance is highly dependent of the bandwidth parameter. \cite{browne2017estimate} proposed a cumbersome algorithm for setting an efficient bandwidth that is not straighforward to implement in practice. As for the estimator of \cite{maume2018estimation}, a large dataset is needed in order to have a low estimation error, as no algorithm of bandwidth parameter selection is established.

To overcome these issues, we explore the random forest algorithm introduced by \citet{breiman2001random} in order to estimate the conditional distribution of the output given an input. The main contribution of this paper is to provide different estimation strategies of the first-order QOSA index based on this method.

The paper is organized as follows. We recall in Section \ref{sec:2:qosa_estimation} the definition of the first-order QOSA index and initiate the estimation process. Section \ref{sec:3:RF} presents the random forest algorithm and several estimators of the first-order QOSA index based on this method are described in Section \ref{sec:4:num_estimator_QOSA}. The entire process is summarized in Section \ref{sec:5:overall_procedure}. Then, the performance of the estimators is investigated in Section \ref{sec:6:num_exp} on simulated data and the relevance of this index is highlighted on a real dataset in Section \ref{sec:7:practical_case}. Finally, a conclusion is given in Section \ref{sec:8:conclusion}.

\section{Estimation of the QOSA index}\label{sec:2:qosa_estimation}

Let us consider the input-output system where $\bX = \left( X_{1}, \dots, X_{d} \right) \in \RR^d$ is a random vector of $d$ independent inputs and $Y = f \left( \bX \right)$ is the output random variable of a measurable deterministic function $f : \RR^d \rightarrow \RR$ which can be a mathematical function or a computational code. Then, given a level $\alpha \in \left] 0,1 \right[$, \citet{fort2016new} introduced the first-order Quantile Oriented Sensitivity Analysis (QOSA) index, related to the input $X_i$, as
\[
S_i^\alpha = \dfrac{\displaystyle \min_{\theta \in \RR} \EE \left[ \psi_\alpha \left( Y,\theta \right) \right] - \EE \left[ \min_{\theta \in \RR} \EE \left[ \left. \psi_\alpha \left( Y,\theta \right)\right| X_i \right] \right]}{\displaystyle \min_{\theta \in \RR} \EE \left[ \psi_\alpha \left(Y,\theta \right) \right]} \ ,
\]
with the contrast function $\psi_\alpha : \left( y,\theta \right) \mapsto \left( y-\theta \right) \left( \alpha - \ind_{\left\lbrace y \leqslant \theta \right\rbrace} \right)$. This function, also called \textit{pinball loss} or \textit{check function} in the literature is the cornerstone of the quantile regression \citep{koenker2001quantile}. Quantile and conditional quantile are related to this loss function as follows
\[
q^\alpha \left( Y \right) = \argmin_{\theta \in \RR} \EE \left[ \psi_\alpha \left(Y,\theta \right) \right] \quad \textnormal{ and } \quad q^\alpha \left( \left. Y \right| X_i \right) = \argmin_{\theta \in \RR} \EE \left[ \left. \psi_\alpha \left( Y,\theta \right) \right| X_i \right] \ ,
\]
where $q^\alpha \left( Y \right)$ is the $\alpha$-quantile of $Y$ and $q^\alpha \left( \left. Y \right| X_i \right)$, the $\alpha$-quantile of $Y$ given $X_i$. Thus, the index $S_i^\alpha$ can be rewritten in the following way,
\[
S_i^\alpha = 1 - \dfrac{\displaystyle \EE \left[ \min_{\theta \in \RR} \EE \left[ \left. \psi_\alpha \left( Y,\theta \right) \right| X_i \right] \right]}{\displaystyle \min_{\theta \in \RR} \EE \left[ \psi_\alpha \left( Y,\theta \right) \right]} = 1 - \dfrac{\EE \left[ \psi_\alpha \left( Y, q^\alpha \left( \left. Y \right| X_i \right) \right) \right]}{\EE \left[ \psi_\alpha \left( Y, q^\alpha \left( Y \right) \right) \right]} = 1 - \dfrac{O}{P} \ ,
\]
where $O$ refers to $\displaystyle \EE \left[ \min_{\theta \in \RR} \EE \left[ \left. \psi_\alpha \left( Y,\theta \right) \right| X_i \right] \right] = \EE \left[ \psi_\alpha \left( Y, q^\alpha \left( \left. Y \right| X_i \right) \right) \right]$ and $P$, to $\displaystyle \min_{\theta \in \RR} \EE \left[ \psi_\alpha \left( Y,\theta \right) \right] = \EE \left[ \psi_\alpha \left( Y, q^\alpha \left( Y \right) \right) \right]$.

Hence, as stated in \cite{browne2017estimate}, the index $S_i^\alpha$ compares the mean distance between $Y$ and its conditional quantile to the mean distance between $Y$ and its quantile, where  the pinball loss function $\psi_\alpha$ is the considered distance. This index has some basic properties requested for a reasonable sensitivity index such as $0 \leqslant S_i^\alpha \leqslant 1$, $S_i^\alpha = 0$ if $Y$ is independent of $X_i$ and $S_i^\alpha =1$ if $Y$ is $X_i$ measurable.

It should be mentioned that \citet{kucherenko2019quantile} proposed new indices \veroc{$K_\alpha$} to assess the impact of inputs on the $\alpha$-quantile of the output distribution. They directly quantify the mean distance between quantiles $q^\alpha \left( Y \right)$ and $q^\alpha \left( \left. Y \right| X_i \right)$ rather than the mean distance between average contrast functions like in the first-order QOSA index. Different estimation strategies are investigated in their paper (brute force Monte Carlo and double-loop reordering approach). But a major limitation is that a large sample size is required to get an accurate computation of the index (samples of size $2^{18}$ are used in their paper). \veroc{Also, as mentioned in \citet{qosa-shapley}, the practical interpretation of the $K_\alpha$ indices is questionable.}

Let us now initiate the estimation procedure for the first-order QOSA index $S_i^\alpha$, associated to a specific input $X_i$ and a level $\alpha$. \\
We consider an i.i.d $n$-sample $\cD_n^\diamond = \left( \bX^{\diamond j}, Y^{\diamond j} \right)_{j=1,\ldots,n}$ such that $Y^{\diamond j} = f \left( \bX^{\diamond j} \right), j=1,\ldots,n$. Then, a first natural estimator of the $P$ term of the QOSA index based on the quantity $\EE \left[ \psi_\alpha \left( Y, q^\alpha \left( Y \right) \right) \right]$ is proposed
\begin{equation}\label{eq:2:denum_estimator_QOSA_v1}
\widehat{P}_1 = \dfrac{1}{n} \sum_{j=1}^n \psi_\alpha \left( Y^{\diamond j}, \widehat{q}^{\alpha} (Y) \right) \ ,
\end{equation}
with $\widehat{q}^{\alpha} (Y)$, the classical empirical estimator for $q^\alpha \left( Y \right)$ obtained from $\cD_n^\diamond$. \\
\veroc{The $P$ term can be alternatively estimated as follows by using the  quantity $\displaystyle \min_{\theta \in \RR} \EE \left[ \psi_\alpha \left( Y,\theta \right) \right]$.}
\[
\widehat{P}_2 = \min_{\theta \in \RR} \dfrac{1}{n} \sum_{j=1}^n \psi_\alpha \left( Y^{\diamond j}, \theta \right) \ ,
\]
where the minimum is reached for one of the elements of $\left( Y^{\diamond j} \right)_{j=1,\ldots,n}$. As the function to minimize is decreasing then increasing, this estimator therefore requires to compute $\frac{1}{n} \sum\limits_{j=1}^n \psi_\alpha \left( Y^{\diamond j}, Y^{\diamond (k)} \right)$, $k=1, \ldots, n$, until it increases, with $Y^{\diamond (k)}$ the order statistics of $\left( Y^{\diamond 1}, \ldots, Y^{\diamond n} \right) $. This process is much more time-consuming than the first estimator where we just need to compute the quantile and then plug it. Thus, in the sequel, we are going to use the $\widehat{P}_1$ estimator.

The $O$ term of the QOSA index is trickier to estimate because a good approximation of the conditional distribution of $Y$ given $X_i$ is necessary. Both existing estimators of the QOSA index currently provided in \citet{browne2017estimate,maume2018estimation} handle this feature thanks to kernel-based methods. But in practice, with these methods,  we are faced with determining the optimal bandwidth parameter or using large sample sizes in order to have a sufficiently low estimation error when employing a non optimal bandwidth. Thus, when dealing with costly computational models, a precise enough  estimation of these indices can be difficult to achieve or even
unfeasible.

We propose in this paper to address these issues by using the random forest method for estimating the conditional distribution. Therefore, several statistical estimators for the $O$ term of the first-order QOSA index will be defined in Section \ref{sec:4:num_estimator_QOSA}. Let us first recall the random forest algorithm.

\section{Random forests}\label{sec:3:RF}

Random forests are ensemble learning methods, first introduced by \citet{breiman2001random}, which can be used in classification or regression problems. We only focus on their use for regression task and assume to be given a training sample $\cD_n = \left( \bX^j, Y^j \right)_{j=1,\ldots,n}$ of i.i.d random variables distributed as the prototype pair $\left( \bX, Y \right)$.

Breiman's forest growns a collection of $k$ regression trees based on the CART procedure described in \citet{breiman1984classification}. Building several different trees from a single dataset requires to randomize the tree building process. Randomness injected in each tree is denoted by $\Theta_\ell$ where $\left( \Theta_\ell \right)_{\ell=1,\ldots,k}$ are independent random variables distributed as $\Theta$ (independent of $\cD_n$). $\Theta = \left( \Theta_1, \Theta_2 \right)$ contains indices of observations selected to build the tree and indices of splitting candidate directions in each cell. 

In more detail, the $\ell$-th tree is built using a bootstrap sample $\cD_n^\star \left( \Theta_\ell \right)$ from the original dataset. Only these observations are used to construct the tree and to make the tree prediction. Once the observations have been selected, the algorithm forms a recursive partitioning of the input space. In each cell, a number $max\mathunderscore features$ of variables is selected uniformly at random among all inputs. Then, the best split is chosen as the one optimizing the CART  splitting criterion only along the $max\mathunderscore features$ preselected directions. This process is repeated in each cell. A stopping criterion, often implemented, is that a split point at any depth will only be considered if it leaves at least $min\mathunderscore samples\mathunderscore leaf$ samples in each of the left and right child nodes. After tree partition has been completed, the prediction of the $\ell$-th tree denoted by $m_n^b \left( \bx;\Theta_\ell,\cD_n \right)$ at a new point $\bx$ is computed by averaging the $N_n^b \left( \bx; \Theta_\ell, \cD_n \right)$ observations falling into the cell $A_n \left( \bx; \Theta_\ell, \cD_n \right)$ of the new point. \\
Hence, the random forest prediction is the average of the $k$ predicted values:
\begin{equation}\label{eq:3:EC_forest_estimator}
m_{k,n}^b \left( \bx;\Theta_1,\ldots,\Theta_k,\cD_n \right) = \dfrac{1}{k}\sum_{\ell=1}^{k} m_n^b \left( \bx;\Theta_\ell,\cD_n \right) = \dfrac{1}{k}\sum_{\ell=1}^{k} \left( \sum_{j \in \cD_n^{\star} \left( \Theta_\ell \right)} \dfrac{\ind_{\left\lbrace \bX^j \in A_n \left( \bx;\Theta_\ell,\cD_n \right) \right\rbrace}}{N_n^b \left( \bx;\Theta_\ell,\cD_n \right)} Y^j \right) \ .
\end{equation}

By defining the random variable $B_{j} \left( \Theta_\ell^1,\cD_n \right)$ as the number of times that the observation $\left( \bX^j, Y^j \right)$ has been used from the original dataset for the $\ell$-th tree construction, the conditional mean estimator in Equation \eqref{eq:3:EC_forest_estimator} is rewritten as follows
\begin{equation}\label{eq:3:bootstrap_weighted_EC_forest_estimator}
m_{k,n}^b \left( \bx;\Theta_1,\ldots,\Theta_k,\cD_n \right) = \sum_{j=1}^{n} w_{n,j}^b \left( \bx; \Theta_1,\ldots,\Theta_k,\cD_n \right) Y^j \ ,
\end{equation}
where the weights $w_{n,j}^b \left( \bx; \Theta_1,\ldots,\Theta_k,\cD_n \right)$ are defined by
\begin{equation}\label{eq:3:bootstrap_weights}
w_{n,j}^b \left( \bx; \Theta_1,\ldots,\Theta_k,\cD_n \right) = \dfrac{1}{k} \sum_{\ell=1}^{k} \dfrac{B_{j} \left( \Theta_\ell^1,\cD_n \right) \ind_{\left\lbrace \bX^j \in A_n \left( \bx;\Theta_\ell,\cD_n \right) \right\rbrace}}{N_n^b \left( \bx;\Theta_\ell,\cD_n \right)} \ .
\end{equation}

A variant of the Equation \eqref{eq:3:bootstrap_weighted_EC_forest_estimator} provides another estimator of the conditional mean. Trees are still grown as in the standard random forest algorithm being based on the bootstrap samples but, for the tree prediction, the original dataset $\cD_n$ is used instead of the bootstrap sample $\cD_n^\star \left( \Theta_\ell \right)$ associated to the $\ell$-th tree and we get
\begin{equation}\label{eq:3:original_weighted_EC_forest_estimator}
m_{k,n}^o \left( \bx;\Theta_1,\ldots,\Theta_k,\cD_n \right) = \sum_{j=1}^{n} w_{n,j}^o \left( \bx; \Theta_1,\ldots,\Theta_k,\cD_n \right) Y^j \ ,
\end{equation}
where the weights $w_{n,j}^o \left( \bx; \Theta_1,\ldots,\Theta_k,\cD_n \right)$ are defined by
\begin{equation}\label{eq:3:original_weights}
w_{n,j}^o \left( \bx; \Theta_1,\ldots,\Theta_k,\cD_n \right) = \dfrac{1}{k} \sum_{\ell=1}^{k} \dfrac{\ind_{\left\lbrace \bX^j \in A_n \left( \bx;\Theta_\ell,\cD_n \right) \right\rbrace}}{N_n^o \left( \bx;\Theta_\ell,\cD_n \right)} \ .
\end{equation}
It has to be noted that contrary to Equation \eqref{eq:3:bootstrap_weights} where $N_n^b \left( \bx;\Theta_\ell,\cD_n \right)$ refers to the number of elements of $\cD_n^\star \left( \Theta_\ell \right)$ falling into $A_n \left( \bx;\Theta_\ell,\cD_n \right)$, in Equation \eqref{eq:3:original_weights}, $N_n^o \left( \bx;\Theta_\ell,\cD_n \right)$ is the number of elements of $\cD_n$ that fall into $A_n \left( \bx;\Theta_\ell,\cD_n \right)$.

Thus, both weighted approaches using, either the bootstrap samples (Equation \eqref{eq:3:bootstrap_weighted_EC_forest_estimator}) or the original dataset (Equation \eqref{eq:3:original_weighted_EC_forest_estimator}), allow to see the random forest method as a local averaging estimate \citep{lin2006random,scornet2016random} and will be at the heart of the strategies proposed for estimating the $O$ term of the QOSA index. In the following, to lighten notation we will omit the dependence to $\Theta$ and $\cD_n$ in the weights.
\section{Estimation of the $O$ term of the QOSA index}\label{sec:4:num_estimator_QOSA}

By using the random forest method aforementioned, ten estimators of the $O$ term may be defined. The first four rely on the expression $\EE \left[ \psi_\alpha \left( Y, q^\alpha \left( \left. Y \right| X_i \right) \right) \right]$ and the others on $\displaystyle \EE \left[ \min_{\theta \in \RR} \EE \left[ \left. \psi_\alpha \left( Y,\theta \right) \right| X_i \right] \right]$. \verob{Since our aim is to estimate conditional expressions with respect to one input variable, say $X_i$, we shall consider forests driven by $X_i$, i.e. the random forest is built with the observations $\cD_n^i = \left( X_i^j, Y^j \right)_{j= 1,\ldots,n}$ from $\cD_n$, which means that $Y$ is explained with $X_i$ only. When needed, we shall denote by $\cD_n^{\star i} $ a bootstrapped sample from $\cD_n^i$ and $\cD_n^{\diamond i} = ( X_i^{\diamond j}, Y^{\diamond j} )_{j= 1,\ldots,n}$ an independant copy of $\cD_n^i$. }

\subsection{Quantile-based $O$ term estimators}\label{sub_sec:4:quantiles_based_qosa}

In this section, the estimations of the $O$ term of the QOSA index are based on the quantity $\EE \left[ \psi_\alpha \left( Y, q^\alpha \left( \left. Y \right| X_i \right) \right) \right]$. Using \verob{two training samples $\cD_n^i$ and $\cD_n^{\diamond i}$}, we define
\[
\widehat{R}_i = \dfrac{1}{n} \sum_{j=1}^n \psi_\alpha \left( Y^{\diamond j}, \widehat{q}^{\alpha} \left( \left. Y \right| X_i = X_i^{\diamond j} \right) \right) \ ,
\]
where the sample \verob{$\cD_n^i$ } is used to get $\widehat{q}^{\alpha} \left( \left. Y \right| X_i = x_i \right)$, an estimator of the conditional quantile $q^{\alpha} \left( \left. Y \right| X_i = x_i \right)$. It is obtained thanks to two approaches based on the random forests, described in the sequel. 

\subsubsection{Quantile estimation with a weighted approach}\label{sub_sub_sec:4:QRF}

\verob{We consider} the estimator of the Conditional Cumulative Distribution Function (C\textunderscore CDF) introduced in \citet{elie2020random} \verob{using $\cD_n^i$ to construct the forest. The C\textunderscore CDF estimator used to estimate the conditional quantile is}
\[
F_{k,n}^b \left( \left. y \right| X_i = x_i \right) = \sum_{j=1}^{n} w_{n,j}^b \left( x_i \right) \ind_{ \{Y^j \leqslant y\}} \ ,
\]
where the $w_{n,j}^b \left( x_i \right)$'s are defined in Equation \eqref{eq:3:bootstrap_weights}.

Hence, given a level $\alpha \in \left[ 0,1 \right]$, the conditional quantile estimator $\widehat{q}^\alpha \left( \left. Y \right| X_i = x_i \right)$ is defined as follows
\[
\widehat{q}^\alpha \left( \left. Y \right| X_i = x_i \right) = \inf_{p=1,\ldots,n} \left\lbrace Y^{p} : F_{k,n}^b \left( \left. Y^p \right| X_i = x_i \right) \geqslant \alpha \right\rbrace \ .
\]
As a result, the estimator of $\EE \left[ \psi_\alpha \left( Y, q^\alpha \left( \left. Y \right| X_i \right) \right) \right]$ based on this method is denoted $\widehat{R}_i^{1, b}$.

Another estimator of the C\textunderscore CDF can be achieved \verob{by  replacing} the weights $w_{n,j}^b \left( x_i \right)$ based on the bootstrap samples of the forest by those using the original dataset $w_{n,j}^o \left( x_i \right)$ provided in Equation \eqref{eq:3:original_weights}. That gives the following estimator which has been proposed in \citet{meinshausen2006quantile},
\[
F_{k,n}^o \left( \left. y \right| X_i = x_i \right) = \sum_{j=1}^{n} w_{n,j}^o \left( x_i \right) \ind_{ \{Y^j \leqslant y\}} \ .
\]
The conditional quantiles are then estimated by plugging $F_{k,n}^o \left( \left. y \right| X_i = x_i \right)$ instead of $F \left( \left. Y \right| X_i = x_i \right)$. Accordingly, the associated estimator of $\EE \left[ \psi_\alpha \left( Y, q^\alpha \left( \left. Y \right| X_i \right) \right) \right]$ based on these weights is denoted $\widehat{R}_i^{1, o}$.

\subsubsection{Quantile estimation within a leaf}\label{sub_sub_sec:4:quantile_in_leaf}

\vero{Let us consider a set of $k$ trees indexed by $\ell=1,\ldots,k$ constructed with the sample $\cD_n^i$.} 

For the $\ell$-th tree, the estimator $\widehat{q}_\ell^{b, \alpha} \left( \left. Y \right| X_i=x_i \right)$ of $q^\alpha \left( \left. Y \right| X_i = x_i \right)$ is obtained with the \vero{bootstrapped} observations \vero{falling into $A_n(x_i;\Theta_\ell, \cD_n^i)$} as follows
\vero{
\[
\begin{split}
\widehat{q}_\ell^{b, \alpha} \left( \left. Y \right| X_i=x_i \right) = \inf_{p=1,\ldots,n} \left\lbrace \right. & Y^{p},\ \left( X_i^p, Y^p \right) \in \cD_n^{i \star}(\Theta_\ell) \textnormal{ and } X_i^p \in A_n(x_i; \Theta_\ell, \cD_n^i): \\
& \left. \sum_{j=1}^n \dfrac{B_{j} \left( \Theta_\ell^1,\cD_n^i \right) \cdot \ind_{\left\lbrace X_i^j\in A_n(x_i;\Theta_\ell,\cD_n^i) \right\rbrace} \cdot \ind_{\left\lbrace Y^j \leqslant Y^p \right\rbrace}}{N_n^b(x_i;\Theta_\ell,\cD_n^i)} \geqslant \alpha \right\rbrace \ .
\end{split}
\]
}
The values from the $k$ randomized trees are then agregated to obtain the following random forest estimate
\[
\verob{\widehat{q}^{b, \alpha}} \left( \left. Y \right| X_i=x_i \right) = \dfrac{1}{k} \sum_{\ell=1}^{k} \widehat{q}_\ell^{b, \alpha} \left( \left. Y \right| X_i=x_i \right) \ .
\]

As for the conditional mean estimate defined in Section \ref{sec:3:RF} or for the C\textunderscore CDF approximation introduced in Subsection \ref{sub_sub_sec:4:QRF}, we can provide a variant using the original sample. Thus, once the forest \vero{is} constructed with the bootstrap samples, \vero{we may estimate the conditional quantiles in the leaves of the $\ell$-th tree using the original sample as follows
\[
\begin{split}
\widehat{q}_\ell^{o, \alpha} \left( \left. Y \right| X_i=x_i \right) = \inf_{p=1,\ldots,n} \left\lbrace \right. & Y^{p},\ \left( X_i^p, Y^p \right) \in \cD_n^i \textnormal{ and } X_i^p \in A_n(x_i; \Theta_\ell, \cD_n^i): \\
& \left. \sum_{j=1}^n \dfrac{\ind_{\left\lbrace X_i^j\in A_n(x_i;\Theta_\ell,\cD_n^i) \right\rbrace} \cdot \ind_{\left\lbrace Y^j \leqslant Y^p \right\rbrace}}{N_n^o(x_i;\Theta_\ell,\cD_n^i)} \geqslant \alpha \right\rbrace \ .
\end{split}
\]
}

%
That gives us the following random forest estimate of the conditional quantile
\[
\verob{\widehat{q}^{o, \alpha}} \left( \left. Y \right| X_i=x_i \right) = \dfrac{1}{k} \sum_{\ell=1}^{k} \widehat{q}_\ell^{o, \alpha} \left( \left. Y \right| X_i=x_i \right) \ .
\]

Thus, these two methods allow us to propose the following estimator $\widehat{R}_i^{2, b}$ (resp. $\widehat{R}_i^{2, o}$) of $\EE \left[ \psi_\alpha \left( Y, q^\alpha \left( \left. Y \right| X_i \right) \right) \right]$ using the bootstrap samples (resp. the original sample).

\subsection{Minimum-based $O$ term estimators}\label{sub_sec:4:minimum_based_qosa}

The estimators developped in Subsection \ref{sub_sec:4:quantiles_based_qosa}, based on $\EE \left[ \psi_\alpha \left( Y, q^\alpha \left( \left. Y \right| X_i \right) \right) \right]$, require to approximate the conditional quantile and then plug it to estimate the $O$ term. As mentioned before, the model $f$ could be time-consuming. Therefore, they may be inappropriate as two training samples are necessary. Hence, we propose in this part to develop estimators of the $O$ term taking advantage from the expression $\displaystyle \EE \left[ \min_{\theta \in \RR} \EE \left[ \left. \psi_\alpha \left( Y,\theta \right) \right| X_i \right] \right]$ for which we \verob{only} need to find the minimum instead of plugging the quantile.

\subsubsection{Minimum estimation with a weighted approach}\label{sub_sub_sec:4:weighted_min}

First of all, a random forest is built with the observations $\cD_n^i$. Then, by considering an additional sample $\left( \bX^{\diamond j} \right)_{j= 1,\ldots,n}$ independent of $\cD_n$, the $O$ term may be estimated as follows
\[
\widehat{Q}_i^{1, b} = \dfrac{1}{n} \sum_{m=1}^n \min_{p=1,\ldots,n} \sum_{j=1}^{n} w_{n,j}^b \left( X_i^{\diamond m} \right) \psi_\alpha \left( Y^j, Y^{p} \right) \ .
\]
Let us notice that the conditional expectation $\EE \left[ \left. \psi_\alpha \left( Y,\theta \right) \right| X_i=x_i \right]$ is estimated with $\sum\limits_{j=1}^{n} w_{n,j}^b \left( x_i \right) \psi_\alpha \left( Y^j, \theta \right)$ \verob{whose  minimum} is reached for $\theta$ equals one of the elements of $\left( Y^j \right)_{j=1,\ldots,n}$.

\verob{Another estimator is obtained by replacing} weigths $w_{n,j}^b \left( x_i \right)$ \verob{ with the} $w_{n,j}^o \left( x_i \right)$ \verob{version} presented in Equation \eqref{eq:3:original_weights} using the original dataset. \verob{The obtained} estimator of the $O$ term \verob{is} denoted by $\widehat{Q}_i^{1, o}$.

\subsubsection{Minimum estimation within a leaf}\label{sub_sub_sec:4:min_in_leaf}

In this subsection, we are going to take advantage of the tree structure in order to propose a new estimator. To begin with, let us consider that a random forest is built with the observations $\cD_n^i$. 

Then, the key point is that an additional sample is no longer required in order to process the outer expectation of the $O$ term. Indeed, for the $\ell$-th tree, the observations falling into its $m$-th leaf node denoted by $A_n \left(m; \Theta_\ell, \cD_n^i \right)$ approximate the conditional distribution of $Y$ given a certain point $X_i=x_i$, which allows to estimate the minimum of the conditional expectation $\displaystyle \min_{\theta \in \RR} \EE \left[ \left. \psi_\alpha \left( Y,\theta \right) \right| X_i=x_i \right]$. Then, we make the average over all the leaves of the $\ell$-th tree to deal with the outer expectation. Hence, \verob{let $N_n^b(m;\Theta_\ell,\cD_n^i)$ be the number of observations of the bootstrap sample $\cD_n^{i \star} \left( \Theta_\ell \right)$ falling into the $m$-th leaf node and  $N_{leaves}^\ell$ be the number of leaves in the $\ell$-th tree. We  define the following tree estimator} for the $O$ term

\begin{align*}
\dfrac{1}{N_{leaves}^\ell} \sum_{m=1}^{N_{leaves}^\ell} \Bigg( \min & \left\lbrace p=1,\ldots,n,\ \left( X_i^p, Y^p \right) \in \cD_n^{i \star}(\Theta_\ell) \textnormal{ and } X_i^p \in A_n \left(m; \Theta_\ell, \cD_n^i \right) \right\rbrace  \\ 
& \left. \sum_{j=1}^n \dfrac{B_{j} \left( \Theta_\ell^1,\cD_n^i \right) \cdot  \psi_\alpha \left( Y^j, Y^{p} \right) \cdot \ind_{\left\lbrace \left( X_i^j, Y^j \right) \in \cD_n^{i \star}(\Theta_\ell),\ X_i^j \in A_n \left(m; \Theta_\ell, \cD_n^i \right) \right\rbrace}}{N_n^b(m;\Theta_\ell,\cD_n^i)} \right) \ .
\end{align*}
The approximations of the $k$ randomized trees are then averaged to obtain the following random forest estimate
\begin{align*}
\widehat{Q}_i^{2, b} = \dfrac{1}{k} \sum_{\ell=1}^k \left[ \dfrac{1}{N_{leaves}^\ell} \sum_{m=1}^{N_{leaves}^\ell} \Bigg( \min \right. & \left\lbrace p=1,\ldots,n,\ \left( X_i^p, Y^p \right) \in \cD_n^{i \star}(\Theta_\ell) \textnormal{ and } X_i^p \in A_n \left(m; \Theta_\ell, \cD_n^i \right) \right\rbrace  \\ 
& \left. \left. \sum_{j=1}^n \dfrac{B_{j} \left( \Theta_\ell^1,\cD_n^i \right) \cdot \psi_\alpha \left( Y^j, Y^{p} \right) \cdot \ind_{\left\lbrace \left( X_i^j, Y^j \right) \in \cD_n^{i \star}(\Theta_\ell),\ X_i^j \in A_n \left(m; \Theta_\ell, \cD_n^i \right) \right\rbrace}}{N_n^b(m;\Theta_\ell,\cD_n^i)} \right) \right] \ .
\end{align*}

As before, the entire procedure described above is still valid using the observations of the original sample falling into the leaves of the $\ell$-th tree instead of the bootstrap ones. Thanks to this change, we get the following estimator of the $O$ term  
\begin{align*}
\widehat{Q}_i^{2, o} = \dfrac{1}{k} \sum_{\ell=1}^k \left[ \dfrac{1}{N_{leaves}^\ell} \sum_{m=1}^{N_{leaves}^\ell} \Bigg( \min \right. & \left\lbrace p=1,\ldots,n,\ \left( X_i^p, Y^p \right) \in \cD_n^i \textnormal{ and } X_i^p \in A_n \left(m; \Theta_\ell, \cD_n^i \right) \right\rbrace  \\ 
& \left. \left. \sum_{j=1}^n \dfrac{\psi_\alpha \left( Y^j, Y^{p} \right) \cdot \ind_{\left\lbrace \left( X_i^j, Y^j \right) \in \cD_n^i,\ X_i^j \in A_n \left(m; \Theta_\ell, \cD_n^i \right) \right\rbrace}}{N_n^o(m;\Theta_\ell,\cD_n^i)} \right) \right] \ .
\end{align*}

It should be noted that looking for the minimum in the leaves directly implies that they are sufficiently sampled for the method to be valid.

\subsubsection{Minimum estimation with a weighted approach and complete trees}\label{sub_sub_sec:4:weighted_min_complete_forest}

In Subsections \ref{sub_sub_sec:4:weighted_min} and \ref{sub_sub_sec:4:min_in_leaf}, the conditional distribution of $Y$ given $X_i$ is obtained from trees grown with $\cD_n^i$. Instead of using this approach, we propose in this part to build a forest with complete trees, i.e. grown with all the model's inputs and then adjust the weights to recover the conditional expectation $\EE \left[ \left. \psi_\alpha \left( Y,\theta \right) \right| X_i \right]$.

Thus, as noticed, a full random forest \verob{is  constructed} with the whole dataset $\cD_n$. Then, by using an additional sample $\left( \bX^{\diamond j} \right)_{j= 1,\ldots,n}$ independent of $\cD_n$, the conditional expectation $\EE \left[ \left. \psi_\alpha \left( Y,\theta \right) \right| X_i=x_i \right]$  is estimated as follows
\begin{align*}
\EE \left[ \left. \psi_\alpha \left( Y,\theta \right) \right| X_i=x_i \right] &= \EE \left[ \left. \EE \left[ \left. \psi_\alpha \left( Y,\theta \right) \right| X_1, \ldots, x_i,\ldots, X_d \right] \right| X_i=x_i \right] \\
&\approx \dfrac{1}{n} \sum_{l=1}^n \left( \sum_{j=1}^n w_{n,j}^{b} \left( \left( \bX_{1}^{\diamond \ell}, \ldots, \bX_{i-1}^{\diamond \ell}, x_i, \bX_{i+1}^{\diamond \ell}, \ldots, \bX_{d}^{\diamond \ell} \right) \right) \right) \\
&\approx \sum_{j=1}^{n} w_{n,j}^{b,i} \left( x_i \right) \psi_\alpha \left( Y^j, \theta \right) \ ,
\end{align*}
where the suitable weights $w_{n,j}^{b,i} \left( x_i \right)$ are defined by
\begin{equation}\label{eq:4:averaged_weights}
w_{n,j}^{b,i} \left( x_i \right) = \dfrac{1}{n} \sum_{\ell=1}^n w_{n,j}^{b} \left( \left( \bX_{-i}^{\diamond \ell}, x_i \right) \right) \ .
\end{equation}
The notation $\bX_{-i}$ indicates the set of all variables except $X_i$ and we note that the conditional expectation given $X_i=x_i$ is recovered by averaging over the components $\bX_{-i}$. Thus, having independent inputs is very convenient. Otherwise, it would be necessary to know the dependency structure in order to generate the observations $\left( \bX_{-i}^{\diamond l} \right)_{l= 1,\ldots,n}$ for each new point $X_i=x_i$, which would make this estimator very cumbersome.

In addition to being used to recover the conditional expectation given $X_i=x_i$,  the sample $\left( \bX^{\diamond j} \right)_{j= 1,\ldots,n}$ is also used to estimate the outer expectation and we finally obtain the following estimator for the $O$ term
\[
\widehat{Q}_i^{3, b} = \dfrac{1}{n} \sum_{m=1}^n \min_{p=1,\ldots,n} \sum_{j=1}^{n} w_{n,j}^{b,i} \left( X_i^{\diamond m} \right) \psi_\alpha \left( Y^j, Y^{p} \right) \ .
\]

By using the weights $w_{n,j}^o \left( \bx \right)$ instead of $w_{n,j}^b \left( \bx \right)$, we may define the estimator $\widehat{Q}_i^{3, o}$.

\section{Overall estimation procedure}\label{sec:5:overall_procedure}

After defining the respective estimators for each term of the first-order QOSA index in Sections \ref{sec:2:qosa_estimation} and \ref{sec:4:num_estimator_QOSA}, the overall estimators are set in the following. In order to improve their accuracy, different strategies are also presented to tune hyperparameters of the random forest.

\subsection{Issues with the leaf size}

When using a random forest method for a regression task, a prediction is generally obtained by using the default values proposed in the packages for the $max\mathunderscore features$ and $min\mathunderscore samples\mathunderscore leaf$ hyperparameters. There are some empirical studies on the impact of these hyperparameters such as \citet{diaz2006gene,scornet2017tuning,duroux2018impact} but no theoretical guarantee to support the default values.

Concerning the estimation methods of the $O$ term of the QOSA index proposed in Section \ref{sec:4:num_estimator_QOSA}, except for $\widehat{Q}_i^{3, b}$ and $\widehat{Q}_i^{3, o}$, it turns out that the values of the hyperparameters must be chosen carefully. \\
First of all, as a forest explaining $Y$ by $X_i$ is built for each model's input, the $max\mathunderscore features$ hyperparameter has no impact in our procedures because it equals 1. Regarding the $min\mathunderscore samples\mathunderscore leaf$ hyperparameter, its impact on the quality of the estimators is investigated through the following toy example
\begin{equation}\label{eq:5:model_diff_expo}
Y = X_1 - X_2 \ ,
\end{equation}
with $X_1, X_2 \sim \cE (1)$. This standard example is commonly used in Sensitivity Analysis literature to assess the quality of QOSA index estimators such as in \citet{fort2016new,browne2017estimate,maume2018estimation}.

\begin{figure}[h]
\makebox[\textwidth][c]{\includegraphics[scale=0.4]{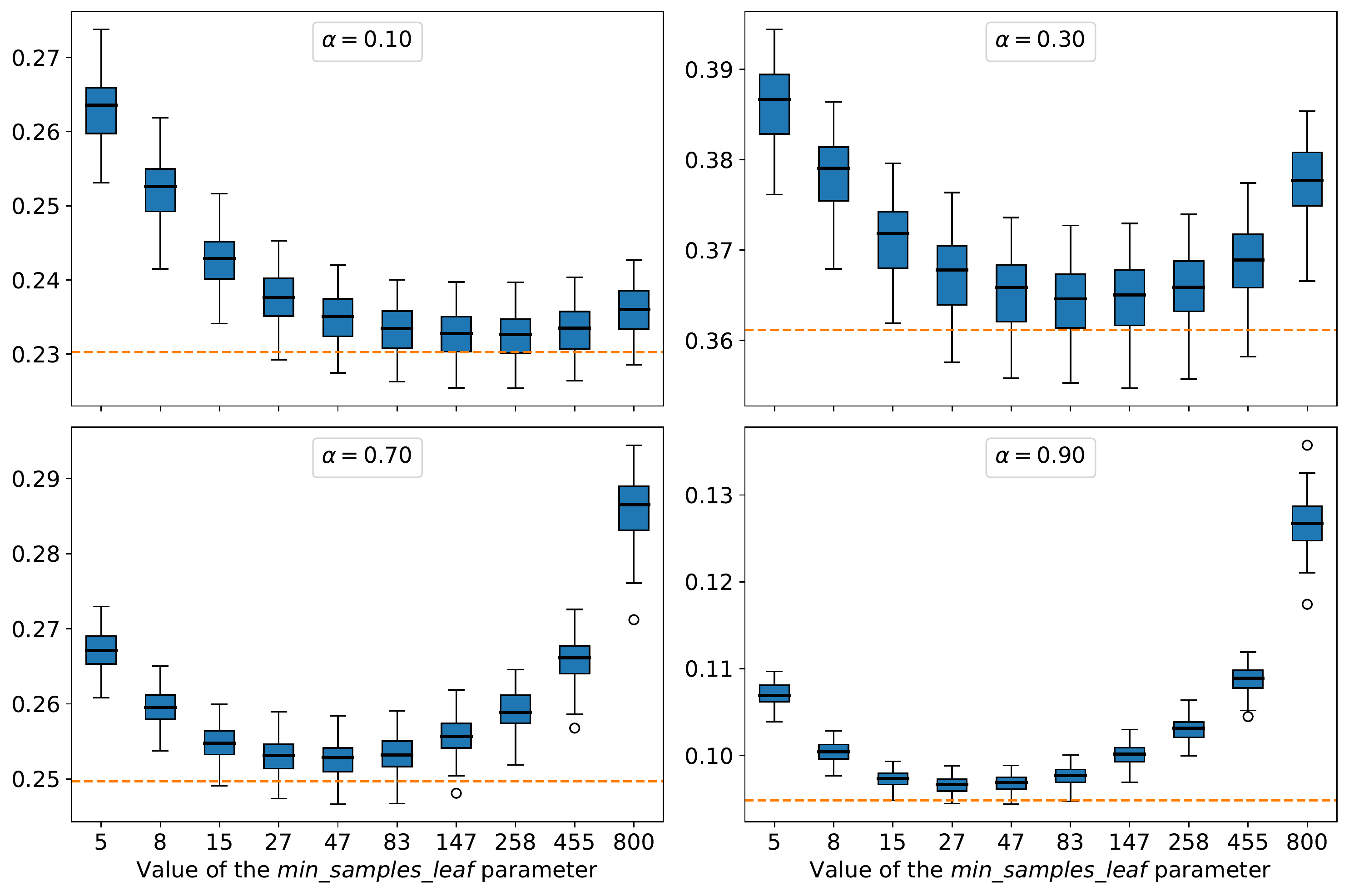}}
\caption{For several levels $\alpha$: distribution of $\widehat{R}_1^{1, o}$, the estimation of the $O$ term associated to the variable $X_1$ for different leaf sizes. The dotted orange line represents the true value on each plot.}
\label{fig:6:distribution_second_term_X1_plug_quant}
\end{figure}

To illustrate the influence of this hyperparameter, we present in Figure \ref{fig:6:distribution_second_term_X1_plug_quant} the boxplot of $\widehat{R}_1^{1, o}$ made with 100 values for different leaf sizes. For each value of $min\mathunderscore samples\mathunderscore leaf$, an estimation $\widehat{R}_1^{1, o}$ is computed using two samples of size $n=10^4$ and a forest grown with $n_{trees} = 500$. Then, the boxplots are compared with the analytical value given below and represented with the dotted orange line on each graph in Figure \ref{fig:6:distribution_second_term_X1_plug_quant}:
\[
\EE \left[ \psi_\alpha \left( Y, q^\alpha \left( \left. Y \right| X_1 \right) \right) \right] = e^{-q^{1 - \alpha} \left( X_2 \right)} \left( 1 + q^{1 - \alpha} \left( X_2 \right) \right) - \alpha \ .
\]

Based on the results obtained in Figure \ref{fig:6:distribution_second_term_X1_plug_quant}, we see that for each level $\alpha$, the performance of $\widehat{R}_1^{1, o}$ depends highly on the choice of the $min\mathunderscore samples\mathunderscore leaf$ hyperparameter. Indeed, with the grid proposed for the values of $min\mathunderscore samples\mathunderscore leaf$, the optimum value seems to be 258 for $\alpha=0.1$, 83 for $\alpha=0.3$, 47 for $\alpha=0.7$ and 27 for $\alpha=0.9$. \\
This issue about the leaf size is only highlighted for $\widehat{R}_i^{1, o}$ but is also encountered for both methods, stated in Subsection \ref{sub_sec:4:quantiles_based_qosa}, computing the conditional quantile with either the bootstrap samples or the original sample.

\begin{figure}[h]
\makebox[\textwidth][c]{\includegraphics[scale=0.4]{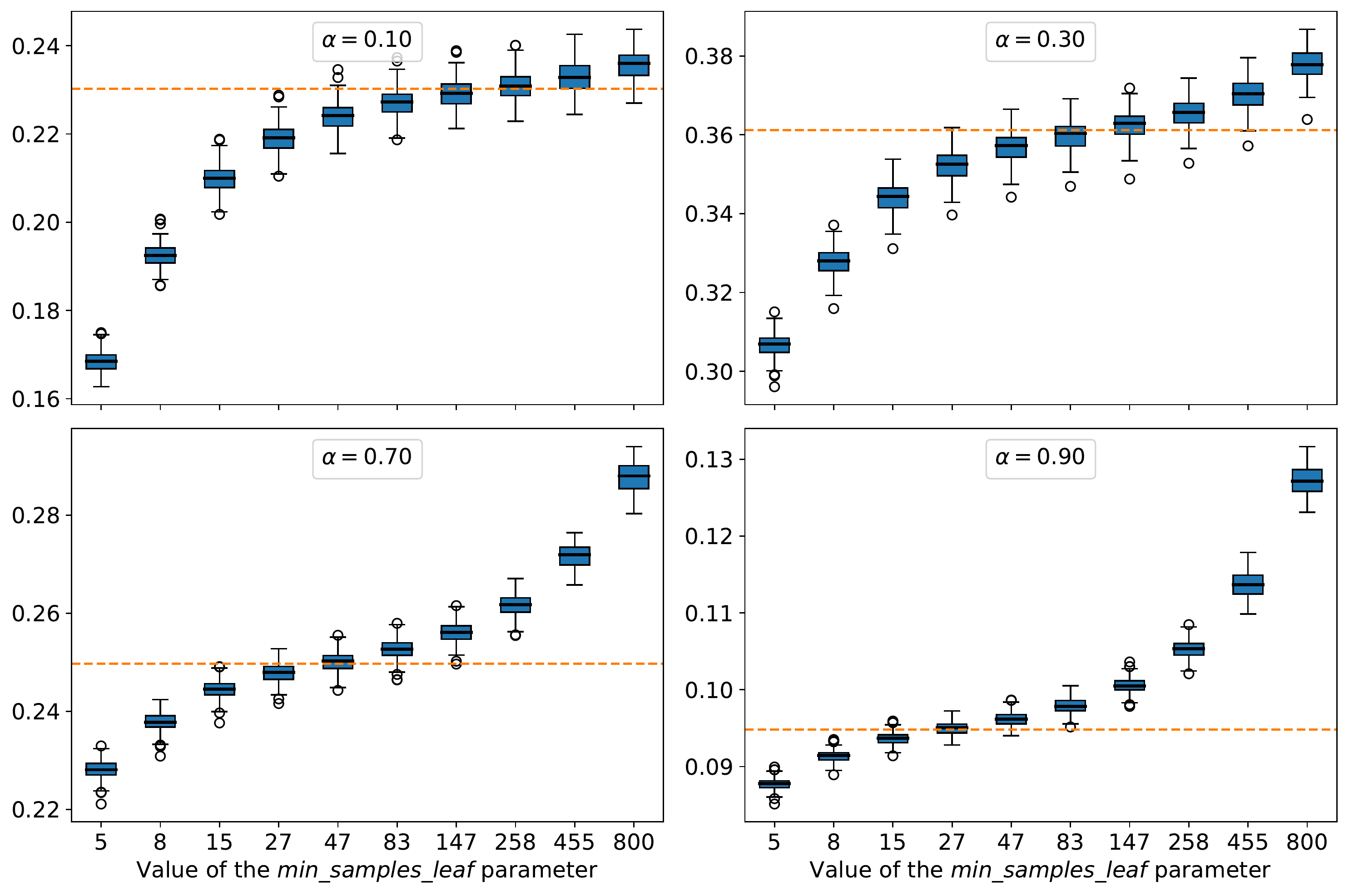}}
\caption{For several levels $\alpha$: distribution of $\widehat{Q}_1^{1, o}$, the estimation of the $O$ term associated to the variable $X_1$ for different leaf sizes. The dotted orange line represents the true value on each plot.}
\label{fig:6:distribution_second_term_X1_min}
\end{figure}
By using the same setting as in Figure \ref{fig:6:distribution_second_term_X1_plug_quant}, the distribution of $\widehat{Q}_1^{1, o}$ is presented in Figure \ref{fig:6:distribution_second_term_X1_min} in order to assess the impact of the $min\mathunderscore samples\mathunderscore leaf$ hyperparameter for a method where the minimum is estimated instead of plugging the quantile. The quality of $\widehat{Q}_1^{1, o}$ also seems to depend on the leaf size and the optimum value, allowing to well estimate $\displaystyle \EE \left[ \min_{\theta \in \RR} \EE \left[ \left. \psi_\alpha \left( Y,\theta \right) \right| X_1 \right] \right]$ for each level $\alpha$, is the same as in Figure \ref{fig:6:distribution_second_term_X1_plug_quant}. \\
As before, this concern about the leaf size was only emphasized for $\widehat{Q}_i^{1, o}$ but is also encountered for both methods, detailed in Subsections \ref{sub_sub_sec:4:weighted_min} and \ref{sub_sub_sec:4:min_in_leaf}, approximating the minimum with either the bootstrap samples or the original sample.

\begin{figure}[!h]
\makebox[\textwidth][c]{\includegraphics[scale=0.4]{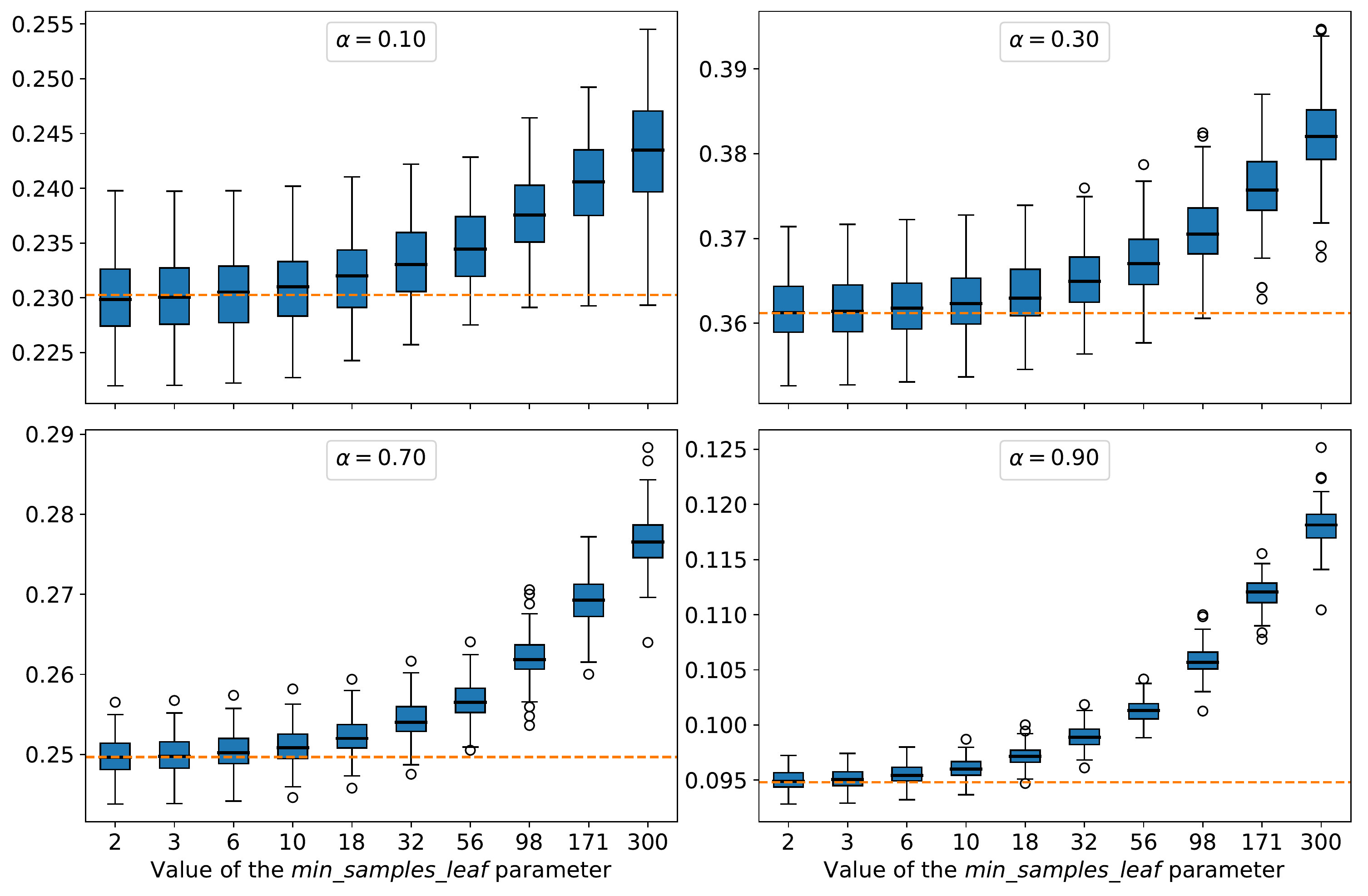}}
\caption{For several levels $\alpha$: distribution of $\widehat{Q}_1^{3, o}$, the estimation of the $O$ term associated to the variable $X_1$ for different leaf sizes. The dotted orange line represents the true value on each plot.}
\label{fig:6:distribution_second_term_X1_min_full_trees}
\end{figure}

For the methods $\widehat{Q}_i^{3, b}$ and $\widehat{Q}_i^{3, o}$, based on complete trees, it seems that the tuning of the leaf size is less important as observed in Figure \ref{fig:6:distribution_second_term_X1_min_full_trees}. Indeed, whatever the $\alpha$ level, the best results are observed for almost fully developed trees.

Thus, for all other estimators of the $O$ term proposed in Section \ref{sec:4:num_estimator_QOSA}, a method giving us the optimal value of the leaf size for each level $\alpha$ is required to properly estimate the first-order QOSA index.

\subsection{Tuning the leaf size}

In order to tune the leaf size of our estimators, two methods are presented in this part. They lead to significatively improve the efficiency of the estimation. The first one rests on a classical cross-validation procedure and the second one uses the Out-Of-Bag samples.

\subsubsection{Cross-validation procedure}

The estimators of the $O$ term developed in Subsection \ref{sub_sec:4:quantiles_based_qosa} are part of the conditional quantile estimation problem. Indeed, in a regression scheme, the conditional mean minimizes the expected squared error loss, while the conditional quantile $q^\alpha \left( \left. Y \right| X_i=x_i \right)$ minimizes the following expected loss
\[
q^\alpha \left( \left. Y \right| X_i \right) = \argmin_{h:\RR \rightarrow \RR} \EE \left[ \psi_\alpha \left( Y, h \left(X_i \right) \right) \right] \ .
\]
Thus, estimators of $\EE \left[ \psi_\alpha \left( Y, q^\alpha \left( \left. Y \right| X_i \right) \right) \right]$ established in Subsection \ref{sub_sec:4:quantiles_based_qosa} allow to assess the quality of the approximation of the true conditional quantile function. The smaller they are, the better the estimate of the conditional quantile function is. That is verified in Figure \ref{fig:6:distribution_second_term_X1_plug_quant} and explains why we have this convex shape depending on the leaf size. As a matter of fact, when the value of the $min\mathunderscore samples\mathunderscore leaf$ hyperparameter is incorrectly chosen, the approximation of the true conditional quantile function is wrong and so, this of $\EE \left[ \psi_\alpha \left( Y, q^\alpha \left( \left. Y \right| X_i \right) \right) \right]$ too. 

Hence, in order to estimate well the conditional quantile function $q^\alpha \left( \left. Y \right| X_i \right)$ and therefore, $\EE \left[ \psi_\alpha \left( Y, q^\alpha \left( \left. Y \right| X_i \right) \right) \right]$ (which is our goal), the optimum value of the leaf size will be chosen within a predefined grid containing potential values as being the one minimizing  the empirical generalization error computed with a $K$-fold cross-validation procedure. A detailed description of this process is given in Algorithm \ref{algo:5:CV_process} with $\widehat{R}_i^{1, o}$ for instance. The principle is the same for all estimators defined in Subsection \ref{sub_sec:4:quantiles_based_qosa}.
\begin{algorithm}[!h]
\DontPrintSemicolon
	\KwIn{
		\begin{itemize}
		\item Datasets: $\cD_n^{\diamond i} = \left( X_i^{\diamond j}, Y^{\diamond j} \right)_{j= 1,\ldots,n}$ from $\cD_n^\diamond$ and $\cD_n^i = \left( X_i^j, Y^j \right)_{j= 1,\ldots,n}$ from $\cD_n$
		\item Number of trees: $k \in \NN^\star$
		\item The order where estimating $\EE \left[ \psi_\alpha \left( Y, q^\alpha \left( \left. Y \right| X_i \right) \right) \right]$ : $\alpha \in \left] 0, 1 \right[$
		\item Grid where looking for the best parameter: $grid \mathunderscore min \mathunderscore samples \mathunderscore leaf$
		\item Number of folds: $K \in \left\lbrace 2,\ldots,n \right\rbrace$
		\end{itemize}
	}
	\KwOut{Estimated value of $\EE \left[ \psi_\alpha \left( Y, q^\alpha \left( \left. Y \right| X_i \right) \right) \right]$ at the $\alpha$-level with $\widehat{R}_i^{1, o}$}
	
	\vspace{\baselineskip}
	\Begin(Cross-validation procedure){
	Randomly split the dataset $\cD_n^i$ into $K$ folds. \;
	\ForEach{$\ell \in grid \mathunderscore min \mathunderscore samples \mathunderscore leaf$}{
	\ForEach{fold}{
		Take the current fold as a test set. \;
		Take the remaining groups as a training set. \;
		Fit a random forest model on the training set with the current $\ell$ as $min \mathunderscore samples \mathunderscore leaf$ hyperparameter. \;
		Evaluate the conditional quantiles at the observations $X_i$ in the test dataset and then compute $\widehat{R}_i^{1, o}$ on the test set. \;
		Retain the estimation obtained. \;
		}
		Summarize the quality related to the current $\ell$ by averaging the $K$ estimated values and save the mean. \;
		}
	}
	Select as optimal value $\ell_{opt}$ for the $min \mathunderscore samples \mathunderscore leaf$ hyperparameter, this one with the smallest mean. \;
	Fit a random forest model on the complete dataset $\cD_n^i$ by fixing the $min \mathunderscore samples \mathunderscore leaf$ hyperparameter to $\ell_{opt}$. \;
	Compute $\widehat{R}_i^{1, o}$ with $\cD_n^{\diamond i}$. \;
	\caption{K-fold cross-validation procedure explained with $\widehat{R}_i^{1, o}$}
	\label{algo:5:CV_process}
\end{algorithm}

It has to be noted that the number of folds $K$ should be chosen carefully. Indeed, a lower value of $K$ results in a more biased estimation of the generalization error, and hence undesirable. In contrast, a larger value of $K$ is less biased, but can suffer from large variability. The choice of $K$ is usually 5 or 10, but there is no formal rule.


\subsubsection{Out-Of-Bag quantile error}\label{sub_sub_sec:5:oob_error_quantile}

The estimators detailed in Subsection \ref{sub_sec:4:quantiles_based_qosa} deserve special attention. Indeed, another less cumbersome approach than cross-validation can be used to tune the leaf size. It is based on an adaptation to our context of the widespread ``Out-Of-Bag'' (OOB) error \citep{breiman1996out} in regression and classification to estimate the generalization error.

$\bullet$ We first adapt the calculation of the OOB error for the conditional quantiles estimated with local averaging estimate of the C\textunderscore CDF proposed in Subsection \ref{sub_sub_sec:4:QRF}. For this purpose, we start by defining the OOB quantile error for $\widehat{R}_i^{1, b}$. \\
Let us fix an observation $(X_i^m, Y^m)$ from $\cD_n^i$ and consider $\cI^m$ as the set of trees built with the bootstrap samples not containing this observation, i.e. for which this one is ``Out-Of-Bag''. The conditional quantile given that $X_i=X_i^m$ is estimated through $F_{k,n}^b \left( \left. Y \right| X_i = x_i \right) = \sum\limits_{j=1}^{n} w_{n,j}^b \left( x_i \right) \ind_{ \{Y^j \leqslant Y\}}$ where the weights are tailored to our context as follows
\[
w_{n,j}^b \left(x_i; \Theta_1,\ldots,\Theta_{\vert \cI^m \vert},\cD_n^i \right) = \dfrac{1}{\vert \cI^m \vert} \sum_{\ell \in \cI^m} \dfrac{B_{j} \left( \Theta_\ell^1,\cD_n^i \right) \ind_{\left\lbrace X_i^j \in A_n \left(x_i;\Theta_\ell,\cD_n^i \right) \right\rbrace}}{N_n^b \left(x_i;\Theta_\ell,\cD_n^i \right)},\ j=1,\ldots,n \ .
\]
Then, $q^\alpha \left( \left. Y \right| X_i=X_i^m \right)$ is estimated by plugging $F_{k,n}^b \left( \left. Y \right| X_i = X_i^m \right)$ instead of $F \left( \left. Y \right| X_i = X_i^m \right)$
\[
\widehat{q}_{oob}^{b, \alpha} \left( \left. Y \right| X_i=X_i^m \right) = \inf_{p=1,\ldots,n} \left\lbrace Y^{p} : F_{k,n}^b \left( \left. Y^p \right| X_i = X_i^m \right) \geqslant \alpha \right\rbrace \ .
\]
After this operation \veroc{is carried out for all}  data in $\cD_n^i$, we calculate the error related to the approximation of the true conditional quantile function, i.e. the empirical generalization error
\[
\widehat{OOB}_i^b = \dfrac{1}{n} \sum_{m=1}^n \psi_\alpha \left( Y^{m}, \widehat{q}_{oob}^{b, \alpha} \left( \left. Y \right| X_i = X_i^{m} \right) \right) \ .
\]
\veroc{We may use the original sample (rather than the bootstrap one) in the definition of the weights:}
%
%

\[
w_{n,j}^o \left(x_i; \Theta_1,\ldots,\Theta_{\vert \cI^m \vert},\cD_n^i \right) = \dfrac{1}{\vert \cI^m \vert} \sum_{\ell \in \cI^m} \dfrac{\ind_{\left\lbrace X_i^j \in A_n \left(x_i;\Theta_\ell,\cD_n^i \right) \right\rbrace}}{N_n^o \left(x_i;\Theta_\ell,\cD_n^i \right)},\ j=1,\ldots,n, j \neq m \ .
\]
\kev{This leads to define $F_{k,n}^o$ as follows
\[
F_{k,n}^o \left( \left. y \right| X_i = x_i \right) = \sum\limits_{\substack{j=1 \\ j \neq m}}^{n} w_{n,j}^o \left( x_i \right) \ind_{ \{Y^j \leqslant y\}} \ .
\]
The estimations of the conditional quantile $\widehat{q}_{oob}^{o, \alpha}$ and of the OOB quantile error $\widehat{OOB}_i^o $ follow.}
  
%
%
%

$\bullet$ Secondly, we adapt the calculation of the OOB error for conditional quantiles estimated directly in tree leaves as introduced in Subsection \ref{sub_sub_sec:4:quantile_in_leaf}. In that sense, define OOB quantile error for $\widehat{R}_i^{2, b}$. \\
Let us fix an observation $(X_i^m, Y^m)$ from $\cD_n^i$ and consider the set of trees built with the bootstrap samples not containing this observation. We then aggregate only the predictions of these trees to make our prediction $\widehat{q}_{oob}^{b, \alpha} \left( \left. Y \right| X_i=X_i^m \right)$ of $q^{\alpha} \left( \left. Y \right| X_i=X_i^m \right)$. After this operation carried out for all the data in $\cD_n^i$, we calculate the error related to the approximation of the true conditional quantile function, i.e. the empirical generalization error
\[
\widehat{OOB}_i^b = \dfrac{1}{n} \sum_{m=1}^n \psi_\alpha \left( Y^{m}, \widehat{q}_{oob}^{b, \alpha} \left( \left. Y \right| X_i = X_i^{m} \right) \right) \ .
\]
\veroc{Again, using the original sample instead of the bootstrap one lead to define $\widehat{OOB}_i^o$.}
%

The advantage of these methods, compared to cross-validation techniques, is that they do not require cutting out the training sample $\cD_n^i$ and take place during the forest construction process.

Thus, given the dataset $\cD_n^i$ and a grid containing potential values of the $min\mathunderscore samples\mathunderscore leaf$ hyperparameter, a random forest is built for each one and the OOB quantile error associated is computed. Then, the optimal hyperparameter is chosen as \veroc{the one with the smallest OOB error}.
 
\subsection{Full estimation procedure}

Now, we have all the components in order to set the estimators of the first-order QOSA index $S_i^\alpha$. These are separated in two classes according to the estimation method adopted for the $O$ term. First of all, with the methods plugging the quantile, we define
\[
\widehat{S}_i^\alpha = 1 - \dfrac{\widehat{R}_i}{\widehat{P}_1} \textnormal{ with } \widehat{R}_i \in \left\lbrace \widehat{R}_i^{1, b}, \widehat{R}_i^{1, o}, \widehat{R}_i^{2, b}, \widehat{R}_i^{2, o} \right\rbrace \ .
\]
The whole procedure integrating the cross-validation process for these methods is detailed in Algorithm \ref{algo:appx:QOSA_estimate_two_samples} (see Appendix \ref{sec:appx:qosa_algorithms}).

On the other hand, regarding the methods based on the minimum to compute the $O$ term, we set
\[
\widehat{S}_i^\alpha = 1 - \dfrac{\widehat{Q}_i}{\widehat{P}_1} \textnormal{ with } \widehat{Q}_i \in \left\lbrace \widehat{Q}_i^{1, b}, \widehat{Q}_i^{1, o}, \widehat{Q}_i^{2, b}, \widehat{Q}_i^{2, o}, \widehat{Q}_i^{3, b}, \widehat{Q}_i^{3, o} \right\rbrace \ .
\]
The estimation process based on the minimum is formalized in Algorithms \ref{algo:appx:QOSA_estimate_weighted_min}, \ref{algo:appx:QOSA_estimate_min_in_leaf} and \ref{algo:appx:QOSA_estimate_weighted_min_and_trees_fully_grown}. For the sake of clarity, they are all gathered in Appendix \ref{sec:appx:qosa_algorithms}
. Algorithm \ref{algo:appx:QOSA_estimate_weighted_min} (resp. \ref{algo:appx:QOSA_estimate_weighted_min_and_trees_fully_grown}) estimating the QOSA index with $\widehat{Q}_i^{1, b}$ or $\widehat{Q}_i^{1, o}$ (resp. $\widehat{Q}_i^{3, b}$ or $\widehat{Q}_i^{3, o}$), needs a full training sample $\cD_n$ as well as a partial one $\left( \bX^{\diamond j} \right)_{j= 1,\ldots,n}$. While estimating the QOSA index with $\widehat{Q}_i^{2, b}$ or $\widehat{Q}_i^{2, o}$ only requires one training sample $\cD_n$. This is a major advantage over methods plugging the quantile that need two full training samples.

So far, no consistency result has been proved for $\widehat{S}_i^\alpha$. These various estimators are reviewed in the next section in order to establish their efficiency in practice. Moreover, all these algorithms are implemented within a \texttt{python} package named \texttt{qosa-indices} available at \citet{qosa_indices}, it can be also freely downloaded on the PyPI website.

\section{Numerical illustrations}\label{sec:6:num_exp}

Let us now carry out some simulations in order to investigate the influence of the hyperparameter optimization algorithm, the impact of the number of trees on our estimators and compare the decrease of the estimation error of each one in function of the train sample-size. From these results, the performance of the two best estimators as well as those based on kernel methods defined in \citet{browne2017estimate,maume2018estimation} is assessed. Then, their scalability is tested on a toy example.

\subsection{Comparison of hyperparameter optimization algorithms}

We start by studying the influence of the hyperparameter optimization algorithm on the performance of our estimators plugging the conditional quantile (i.e. using $\widehat{R}_i^{1, b}, \widehat{R}_i^{1, o}, \widehat{R}_i^{2, b}$ or $\widehat{R}_i^{2, o}$). This survey is carried out with the model introduced in Equation \eqref{eq:5:model_diff_expo} and the following setting.

The estimators of the QOSA index are computed with samples of size $n=10^4$. The leaf size is tuned for each estimator over a grid with 20 numbers evenly spaced ranging from 5 to 300 by using either the strategy based on the OOB quantile error developed in Subsection \ref{sub_sub_sec:5:oob_error_quantile} or a $3$-fold cross-validation procedure. Then, to assess the efficiency of each method (CV vs OOB), the experiment is repeated $s=100$ times and the following metrics are computed
\begin{align}\label{eq:6:metrics}
& RMSE_i^\alpha = \sqrt{\dfrac{1}{s} \sum_{j=1}^s \left( \widehat{S}_i^{\alpha, j} - S_i^\alpha \right)^2} \ , \nonumber \\
& Bias_i^\alpha  = \left| \dfrac{1}{s} \sum_{j=1}^s \widehat{S}_i^{\alpha, j} - S_i^\alpha \right| \ , \\
& Variance_i^\alpha  = \dfrac{1}{s} \sum_{j=1}^s \left( \widehat{S}_i^{\alpha, j} - \dfrac{1}{s} \sum_{j=1}^s \widehat{S}_i^{\alpha, j} \right)^2 \ , \nonumber
\end{align}
with $S_i^\alpha$, the analytical values that were provided in \citet{fort2016new}.

\begin{figure}[h!]
	\centering
	\makebox[\textwidth][c]{\includegraphics[width=1.\textwidth]{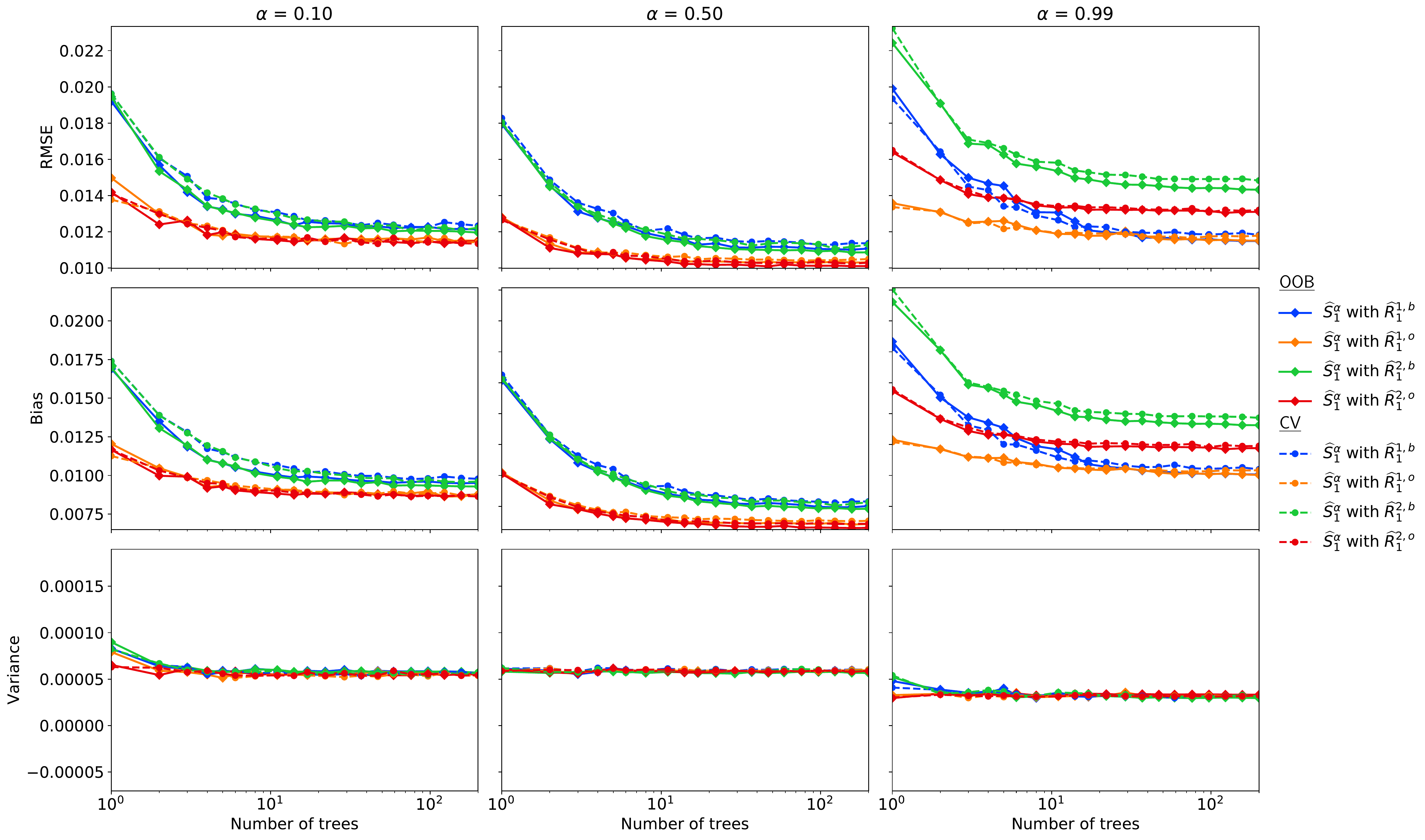}}
	\caption{Evolution of RMSE, bias and variance of the estimators associated with $X_1$, calculated with either the OOB strategy or the Cross-Validation procedure, in function of the number of trees for three levels $\alpha$.}
	\label{fig:6:Cv_vs_OOB}
\end{figure}
In Figure \ref{fig:6:Cv_vs_OOB}, for three levels $\alpha$, we present the evolution of the different metrics related to the variable $X_1$ of our toy example in function of the number of trees ranging from $1$ to $200$ (in log scale). More precisely, \veroc{sub-figures at the top of Figure \ref{fig:6:Cv_vs_OOB}} show the Root Mean Square Error (RMSE), in the middle, the bias and the variance at the bottom.

We observe that regardless of the level $\alpha$ and the number of trees, our estimators plugging the quantile have globally the same performance when calculated with either the OOB strategy or the cross-validation procedure. But, the run time is faster when using the OOB strategy rather than the cross-validation procedure.

\subsection{Convergence with the number of trees and the train sample-size}

We analyze in this part the impact of the number of trees on the performance of all our estimators except for those using $\widehat{Q}_i^{3, b}$ and $\widehat{Q}_i^{3, o}$ because of the computational cost. This survey is also carried out with the model introduced in Equation \eqref{eq:5:model_diff_expo} and the following setting.

The estimators of the QOSA index are computed with samples of size $n=10^4$. The leaf size is tuned over a grid with 20 numbers evenly spaced ranging from 5 to 300 by using a $3$-fold cross-validation procedure for $\widehat{R}_i^{1, b}$ and $\widehat{R}_i^{1, o}$ while the strategy based on the OOB samples,  developed in Subsection \ref{sub_sub_sec:5:oob_error_quantile}, is used for $\widehat{R}_i^{2, b}$ and $\widehat{R}_i^{2, o}$. Regarding the minimum based estimators, the optimal leaf size is obtained via $\widehat{R}_i^{1, o}$ during the $3$-fold cross-validation process. Then, the efficiency of our estimators is assessed with the metrics introduced in Equation \eqref{eq:6:metrics} by repeating the experiment $s=200$.

\begin{figure}[h!]
	\centering
	\makebox[\textwidth][c]{\includegraphics[width=1.\textwidth]{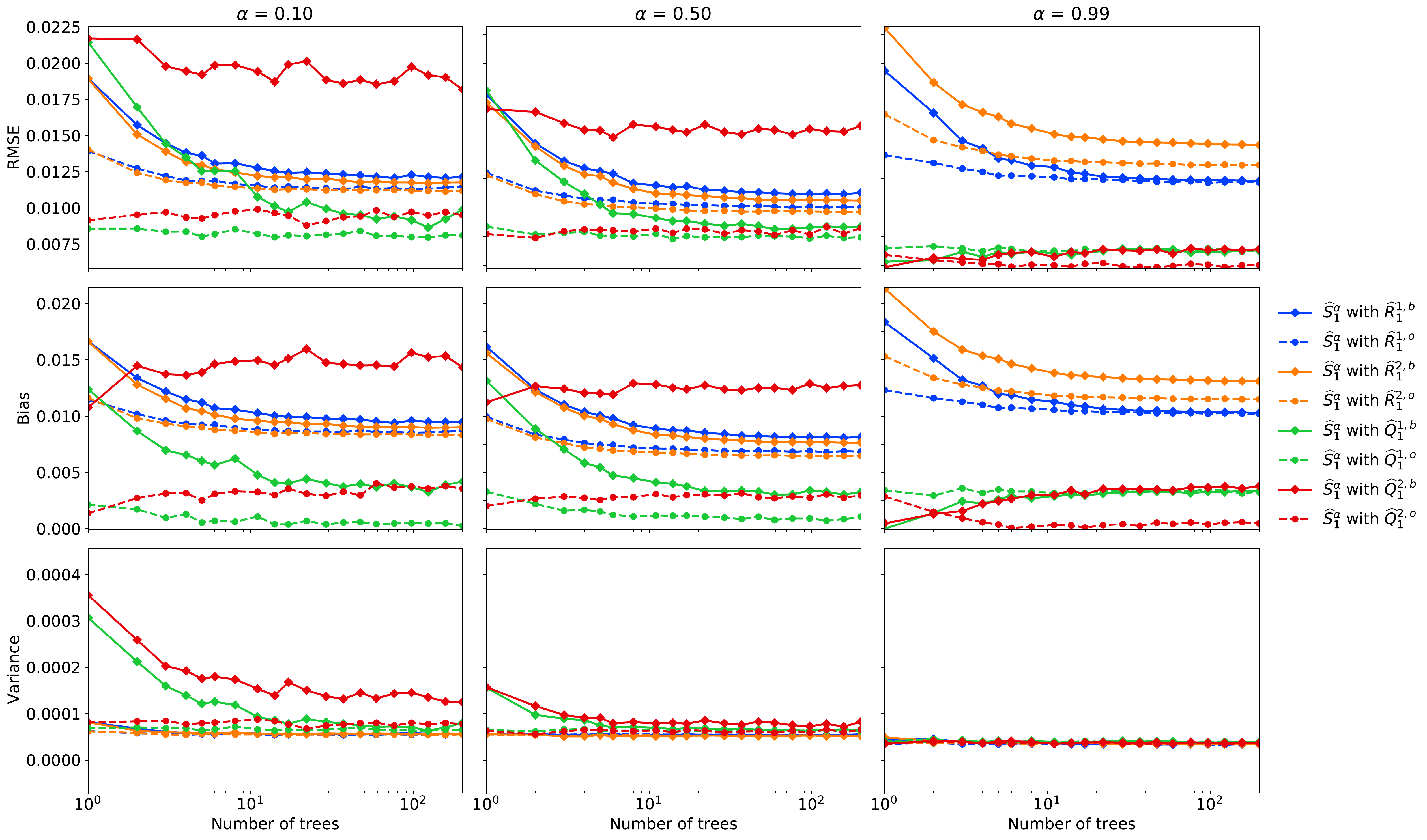}}
	\caption{Evolution of RMSE, bias and variance of the estimators associated with $X_1$ in function of the number of trees for three levels $\alpha$.}
	\label{fig:6:impact_of_trees}
\end{figure}
In Figure \ref{fig:6:impact_of_trees}, for three levels $\alpha$, we present the evolution of the different metrics related to the variable $X_1$ of our toy example in function of the number of trees ranging from $1$ to $200$ (in log scale). More precisely, sub-figures at the top of Figure \ref{fig:6:impact_of_trees} show the Root Mean Square Error (RMSE), in the middle, the bias and the variance at the bottom.

We observe that regardless of the level $\alpha$, RMSE of our estimators is small. The number of trees seems to have no impact for those using $\widehat{Q}_i^{1, o}$ and $\widehat{Q}_i^{2, o}$ as the RMSE value is almost always the same. RMSE of the others decreases in function of the number of trees until it reaches a threshold starting at about 50 trees. Indeed, it is well known that from a certain number, increasing the number of trees becomes useless but results in higher calculation costs. However, we did not expect to have a stable estimation error with so few trees. \\
Besides, still from the RMSE curves, it first appears that the estimators using the original sample (plain lines) have a lower error compared to those using the bootstrap samples (dotted lines). On the other hand, the performance of the minimum based estimators (green and red lines) seems better than those based on the quantile (blue and orange lines). That might be explained by the additional error due to the estimation of the conditional quantile. \\
Variance of all estimators is close to 0 and the bias curves have the same behavior as RMSE curves. This means that bias is the main/only source of error in the RMSE. This bias could be reduced by taking a larger grid where looking for the optimal leaf size during the cross-validation or using another more efficient method to find the optimum.

Let us now compare the decrease of the estimation error in function of the train sample-size. 
As observed in Figure \ref{fig:6:impact_of_trees}, take a very large number of trees is not required in order to have a stable estimation error. Thus, we take $n_{trees} = 100$ and the same setting as before for other parameters in the next study and observe the evolution of the metrics introduced in Equation \eqref{eq:6:metrics} in function of the sample size.

\begin{figure}[h]
	\centering
	\makebox[\textwidth][c]{\includegraphics[width=1.\textwidth]{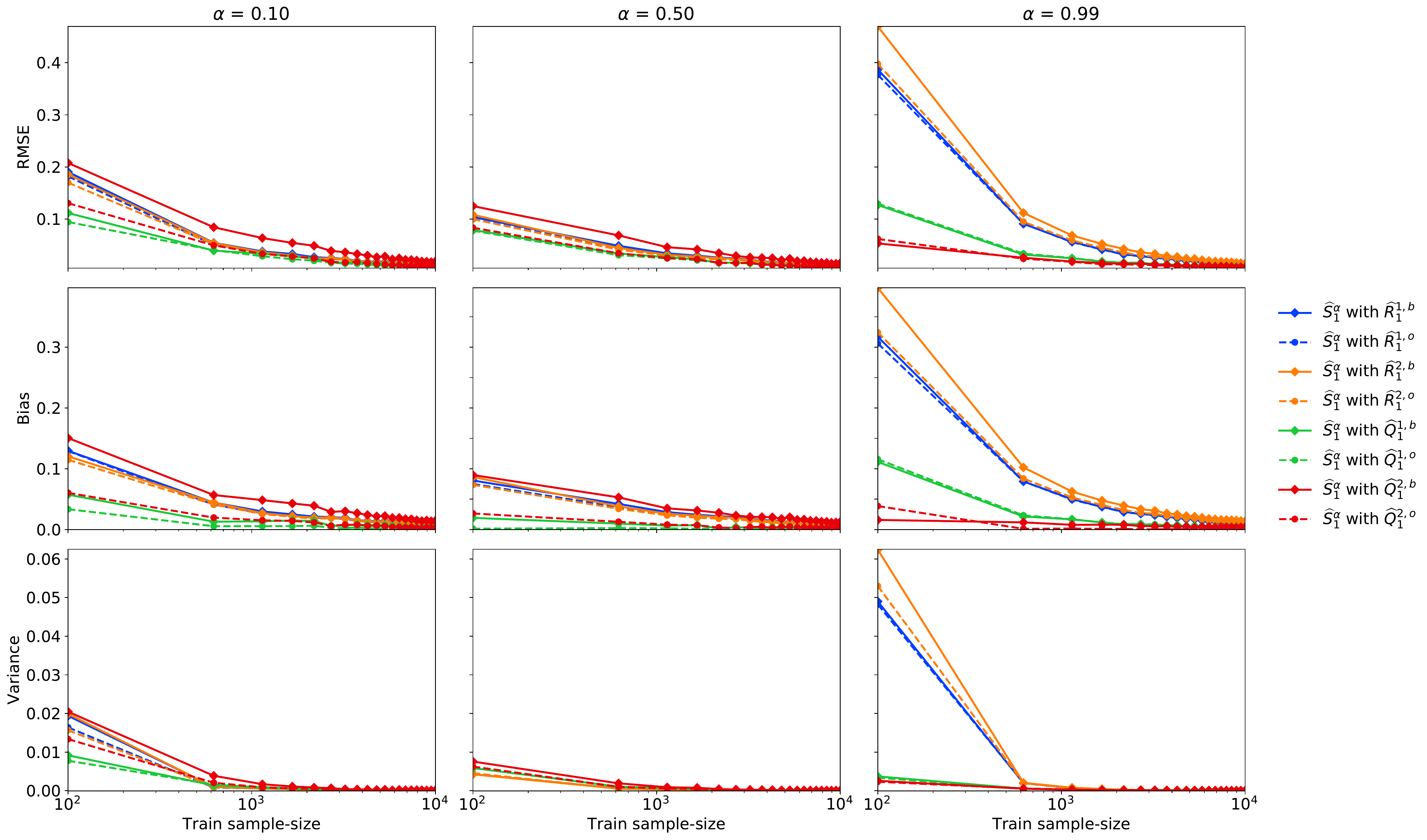}}
	\caption{Evolution of RMSE, bias and variance of the estimators associated with $X_1$ in function of the train sample-size for three levels $\alpha$.}
	\label{fig:6:impact_of_size_sample}
\end{figure}
Figure \ref{fig:6:impact_of_size_sample} presents RMSE, bias and variance of our estimators for different sample sizes. We observe that all the metrics associated with the various estimators converge to 0 at different rates. Indeed, the convergence rates of the metrics of the quantile-based estimators are slower than those based on the minimum.

Hence, from our experiments, it turns out that the minimum-based estimators give the best results. This is an interesting feature because they need less data than those plugging the quantile. Furthermore, few trees are necessary in order to reduce the estimation error. It therefore allows to get a good estimation of the indices with a reasonable computational cost.

\subsection{Comparison with kernel methods}

In this subsection, we compare on the toy example introduced in Equation \eqref{eq:5:model_diff_expo}:
\begin{itemize}
\item the kernel-based estimators proposed in \citet{browne2017estimate,maume2018estimation} denoted by $\check{S}_{i}^{\alpha}$ and $\widetilde{S}_{i}^{\alpha}$,
\item the minimum-based QOSA index estimators building one forest for each input and using the original sample,
\item and the minimum-based QOSA index estimators using a forest grown with trees fully developed.
\end{itemize}

The estimators of the QOSA indices are computed with samples of size $n=10^4$. \\
Forest methods are grown with $n_{trees}=100$. The optimal leaf size for the minimum-based estimators building one forest for each input is obtained with $\widehat{R}_i^{1, o}$ during the 3-fold cross-validation process over a grid containing 20 numbers evenly spaced ranging from 5 to 300. Regarding the minimum-based estimators using a forest grown with trees fully developed, the $min\mathunderscore samples \mathunderscore leaf$ hyperparemeter equals 2. \\
In order to have comparable methods, a cross-validation procedure is also implemented for the kernel-based estimators to choose the optimal bandwidth parameter. It is selected within over a grid containing 20 potential values ranging from 0.001 to 1. Then, we assess the performance of the different estimators by computing their empirical root mean squared error with 100 experiments.

\begin{table}[h]
\centering
\begin{adjustbox}{max width=1.3\textwidth,center}
\begin{tabular}{C{1.5cm} I C{0.9cm} | C{0.9cm} I C{0.9cm} | C{0.9cm} I C{0.9cm} | C{0.9cm} I C{0.9cm} | C{0.9cm} I C{0.9cm} | C{0.9cm} I C{0.9cm} | C{0.9cm}}
\hline
\multirow{2}{*}{} & \multicolumn{2}{cI}{$\widehat{S}_{i}^{\alpha}$ with $\widehat{Q}_{i}^{1,o}$} & \multicolumn{2}{cI}{$\widehat{S}_{i}^{\alpha}$ with $\widehat{Q}_{i}^{2,o}$} & \multicolumn{2}{cI}{$\widehat{S}_{i}^{\alpha}$ with $\widehat{Q}_{i}^{3,b}$} & \multicolumn{2}{cI}{$\widehat{S}_{i}^{\alpha}$ with $\widehat{Q}_{i}^{3,o}$} & \multicolumn{2}{cI}{$\check{S}_{i}^{\alpha}$} & \multicolumn{2}{c}{$\widetilde{S}_{i}^{\alpha}$} \\ \cline{2-13}
& $X_1$ & $X_2$ & $X_1$ & $X_2$ & $X_1$ & $X_2$ & $X_1$ & $X_2$ & $X_1$ & $X_2$ & $X_1$ & $X_2$ \\ \hline
$\alpha = 0.1$ & 0.007 & 0.006 & 0.009 & 0.006 & 0.017 & 0.006 & 0.017 & 0.006 & 0.020 & 0.044 & 0.061 & 0.006 \\ \hline
$\alpha = 0.25$ & 0.008 & 0.006 & 0.009 & 0.006 & 0.013 & 0.007 & 0.013 & 0.007 & 0.013 & 0.036 & 0.042 & 0.012 \\ \hline
$\alpha = 0.5$ & 0.008 & 0.006 & 0.008 & 0.007 & 0.010 & 0.009 & 0.010 & 0.009 & 0.019 & 0.021 & 0.027 & 0.025 \\ \hline
$\alpha = 0.75$ & 0.008 & 0.007 & 0.008 & 0.008 & 0.008 & 0.014 & 0.008 & 0.014 & 0.035 & 0.012 & 0.014 & 0.042 \\ \hline
$\alpha = 0.99$ & 0.006 & 0.016 & 0.006 & 0.018 & 0.006 & 0.032 & 0.006 & 0.032 & 0.084 & 0.071 & 0.013 & 0.11 \\ \hline
& \multicolumn{2}{cI}{} & \multicolumn{2}{cI}{} & \multicolumn{2}{cI}{} & \multicolumn{2}{cI}{} & \multicolumn{2}{cI}{} & \multicolumn{2}{c}{} \\ \hline
run time & \multicolumn{2}{cI}{1 hr} & \multicolumn{2}{cI}{18 min 24 sec} & \multicolumn{2}{cI}{10 hr 41 min} & \multicolumn{2}{cI}{8 hr 18 min} & \multicolumn{2}{cI}{1 hr 55 min} & \multicolumn{2}{c}{1 min 51 sec} \\ \hline
\end{tabular}
\end{adjustbox}
\caption{RMSE and run time for the toy example of the random forest based estimators: $\widehat{S}_{i}^{\alpha}$ computed with $\widehat{Q}_{i}^{1,o}$, $\widehat{Q}_{i}^{2,o}$, $\widehat{Q}_{i}^{3,b}$ and $\widehat{Q}_{i}^{3,o}$ as well as those based on kernel: $\widetilde{S}_{i}^{\alpha}$ and $\check{S}_{i}^{\alpha}$.}
\label{tab:6:comparison_forest_kernel}
\end{table}

Table \ref{tab:6:comparison_forest_kernel} contains the empirical root mean square error of the different estimators associated \veroc{to each} input as well as the overall run time requested to obtain them. About their performance, it seems that the random forest-based estimators are better than the kernel methods. Nevertheless, as regards the methods using $\widehat{Q}_{i}^{3,b}$ and $\widehat{Q}_{i}^{3,o}$, while they have a low error and do not need to tune the leaf size, their run time with the current implementation is too long to be used in practice. Accordingly, we recommend to compute the indices with $\widehat{Q}_{i}^{1,o}$ and $\widehat{Q}_{i}^{2,o}$ in order to get good estimations of the first-order QOSA indices in a reasonable time.

\subsection{Scalability of the methods}

The influence of the model's dimension $d$ over the performance of the estimators using $\widehat{Q}_{i}^{1,o}$ and $\widehat{Q}_{i}^{2,o}$ is investigated in this subsection with the following additive \veroc{exponential framework}
\begin{equation}
Y = \sum_{i=1}^d X_i \ .
\end{equation}
Independent inputs $X_i, i =1, \ldots,d$, follow an Exponential distribution $\cE ( \lambda_i )$, \veroc{with distinct $\lambda_i$.  The} resulting output $Y$ is a generalized Erlang distribution also called Hypoexponential distribution. By taking advantage of the other expression of the first-order QOSA index given in \citet{maume2018estimation}, we obtain the following semi closed-form analytical formula
\begin{equation}\label{eq:6:qosa_index_hypo_distrib}
S_i^\alpha = 1 - \dfrac{\alpha \EE \left[ Xs_{(-i)} \right] - \EE \left[ Xs_{(-i)} \ind_{\left\lbrace Xs_{(-i)} \leqslant q^\alpha \left( Xs_{(-i)} \right) \right\rbrace} \right]}{\alpha \EE \left[ Y \right] - \EE \left[ Y \ind_{\left\lbrace Y \leqslant q^\alpha (Y) \right\rbrace} \right]} \ ,
\end{equation}
with $Xs_{(-i)} = \sum\limits_{j \neq i} X_j$ that also follows a Hypoexponential distribution. Knowing the cumulative distribution function of the Hypoexponential distribution, quantiles $q^\alpha (Y)$ and $q^\alpha \left( Xs_{(-i)} \right)$ are computed by numeric inversion and the analytical expression of the truncated expectations is derived from \citet{marceau2013modelisation}.

For a specific dimension $d$, $d$ values evenly spaced are selected from the interval $\left[ 0.3, 1.25 \right]$ and then each one represents the $\lambda_i$ parameter of an input $X_i, \ i=1, \ldots, d$. QOSA index estimations are then computed with samples of size $n=10^4$, a forest grown with $n_{trees}=100$ and the setting defined hereafter. The leaf size is tuned with $\widehat{R}_i^{1, o}$ over a grid with 20 numbers evenly spaced ranging from 5 to 300 by using a $3$-fold cross-validation. Each experiment is done 100 times in order to compute the RMSE defined in Equation \eqref{eq:6:metrics} for each input, and then we take the weighted mean by the analytical values of the QOSA indices over all dimensions in order to get a global measure.

\begin{figure}[h]
	\centering
	\makebox[\textwidth][c]{\includegraphics[width=1.\textwidth]{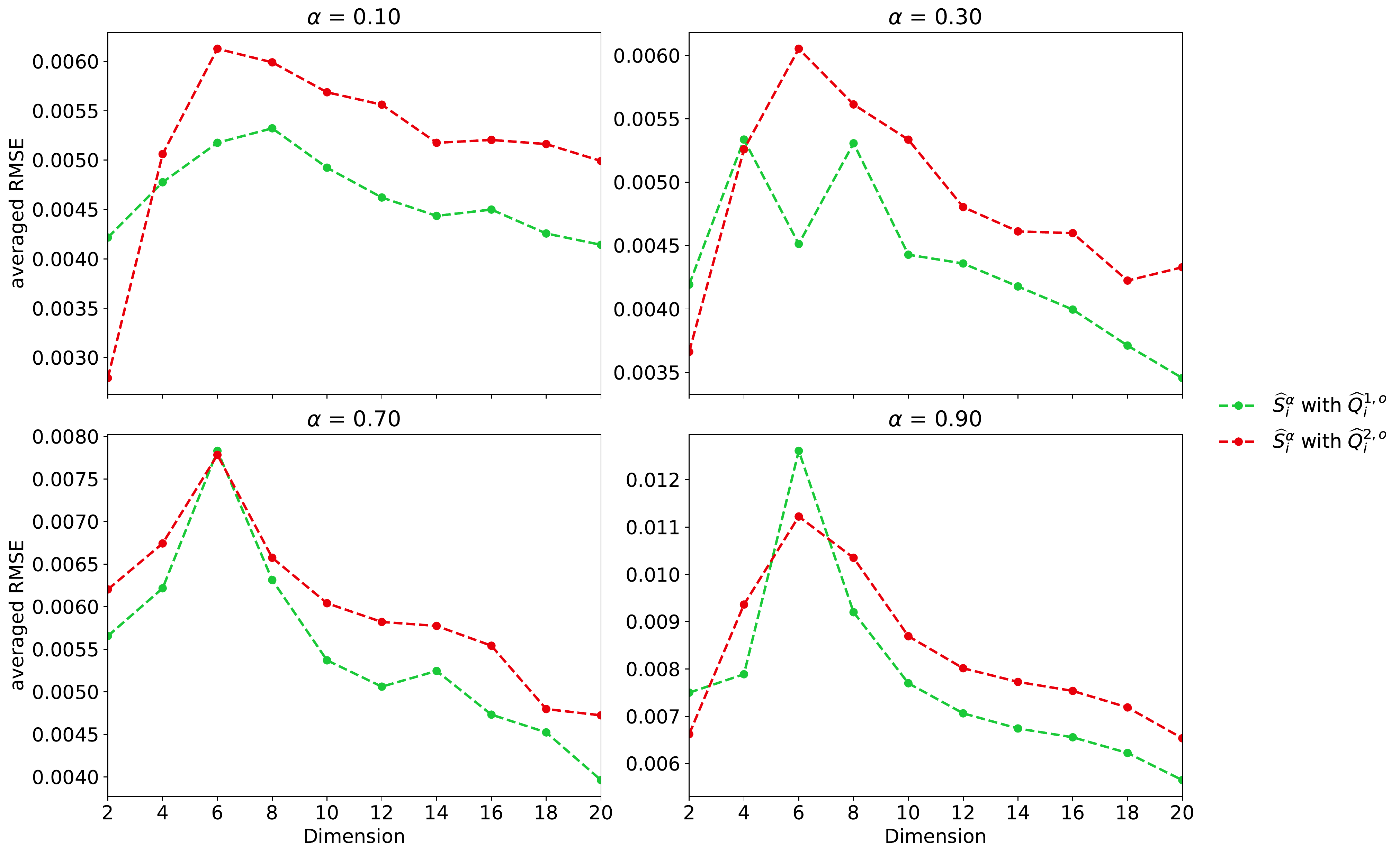}}
	\caption{Evolution of the averaged RMSE over all dimensions of the estimators calculated with $\widehat{Q}_{i}^{1,o}$ and $\widehat{Q}_{i}^{2,o}$ in function of the model dimension for four levels $\alpha$.}
	\label{fig:6:impact_dimension}
\end{figure}
Figure \ref{fig:6:impact_dimension} presents the weighted RMSE as a function of the increasing dimension of our model for several levels $\alpha$. For each one, we observe that the error increases slowly at the beginning until the dimension 6 for both methods then decreases. This phenomenon is due to the chosen parametrization. Indeed, when increasing the dimension of the model, the respective impact of each input is reduced. Thus, from a certain dimension, all the analytical values of the first-order QOSA indices become small and even close to 0 for some inputs. Our estimators properly capture this trend as they decrease by increasing the dimension. However, the estimator using  $\widehat{Q}_{i}^{1,o}$ seems better than this one employing $\widehat{Q}_{2}^{1,o}$ as its error is lower.

\section{Practical case study}\label{sec:7:practical_case}

We propose to apply our methodology to a practical case study. It concerns bias between the predictions from MOCAGE (Mod{\`e}le de Chimie Atmosph{\'e}rique {\`a} Grande Echelle)\footnote{Large Scale Atmospherical Chemestrial Model} and the observed ozone concentration.

This dataset has been proposed and studied in \citet{besse2007comparaison}, the full data description may be found there. It contains $10$ variables with $1041$ observations.  \textbf{O3obs}: observed ozone concentration will be explained by the $9$ other variables described below.
\begin{center}
\begin{tabular}{p{7.5cm}|p{7.5cm}}
\hline
JOUR: type of day ($0$ for holiday vs $1$ for non holiday) & STATION: site of observations (5 different sites)  \\
RMH2O: humidity ratio  &  NO2: nitrogen dioxide concentration \\
VentMOD: wind force & VentANG: wind direction  \\
NO: nitric oxide concentration &  TEMPE: officially predicted temperatures\\
MOCAGE: ozone concentration predicted by a fluid mechanics model &  \\
\hline
\end{tabular}
\end{center}

\vspace{\baselineskip}

Figure \ref{fig:7:qosa_indices_mocage} below gives the QOSA estimations, using the $\widehat{Q}_i^{2, o}$ estimator, since from our numerical study, it is the quicker, requires only one sample and is efficient. The inputs are ranked, for different values of alpha. On the right picture, the QOSA indices are in percentage (normalised by the sum of QOSA indices for all variables).

\begin{figure}[h]
	\centering
	\makebox[\textwidth][c]{\includegraphics[scale=0.45]{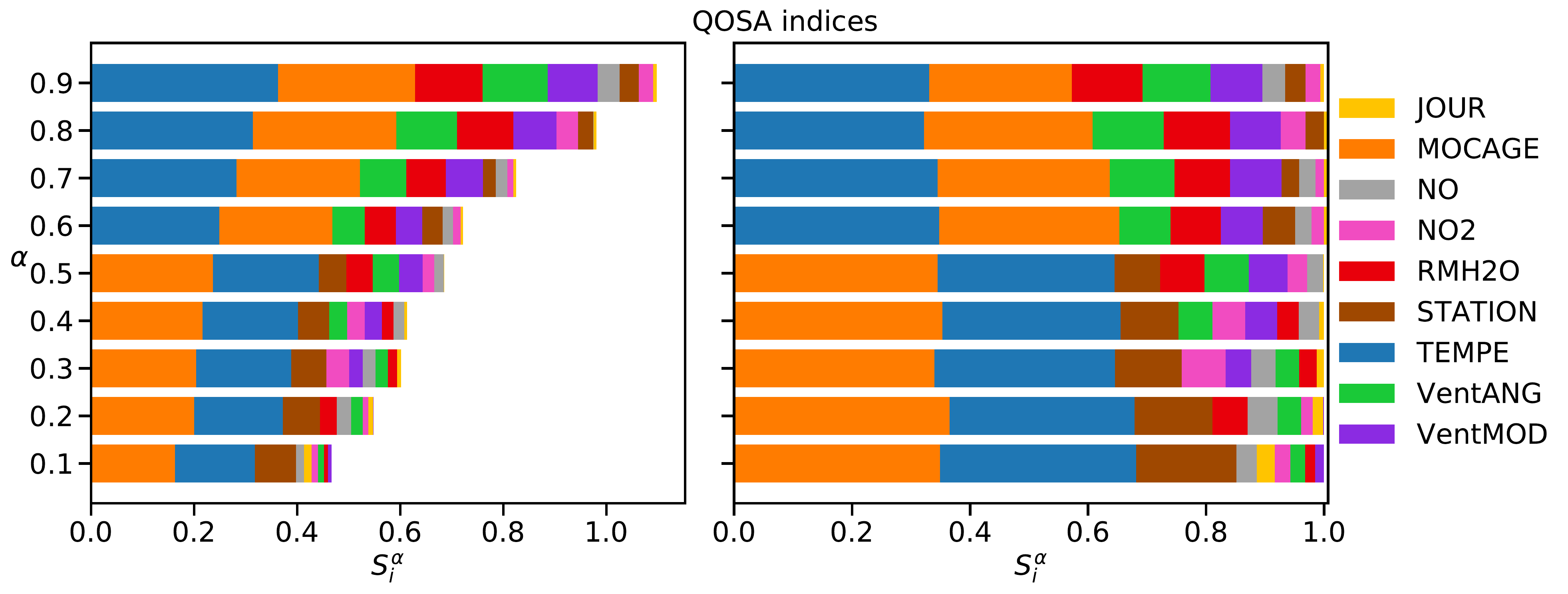}}
	\caption{QOSA (resp. normalised QOSA) indices at different levels $\alpha$ on the left-hand (resp. on the right-hand) plot.}
	\label{fig:7:qosa_indices_mocage}
\end{figure}

In \cite{besse2007comparaison,broto2020variance} the central effects have been studied and lead to consider MOCAGE and TEMPE as the most influencial variables, then STATION and NO2. Our study considers various quantile level effects and shows that for quantile levels greater than 0.6, wind and RMH2O are also important, more than STATION. 

\section{Conclusion}\label{sec:8:conclusion}

In this paper, we introduced several estimators for the first-order QOSA index by using the random forest method. Some of them use the original sample while the others use the bootstrap samples generated during the forest construction. Both classes of estimator seem to be efficient even if we observe in our experiments that the methods using the original sample have a lower estimation error than those \veroc{based on} the bootstrap ones. Thus, supplementary studies should be conducted to inquire into this difference. Furthermore, the performance of these methods is highly dependent on the leaf size. This parameter could be compared to the bandwidth parameter of kernel estimators as it controls the bias of the method. But, it turns out to be easier to calibrate and we propose two methods to do this\veroc{: $K$-fold cross-validation and Out of Bag samples based selection method}. 

It is also well known for \veroc{random forest methods} that the number of trees $k$ should be chosen large enough to reach the desired statistical precision and small enough to make the calculations feasible as the computational cost increases linearly with $k$ as mentioned in \citet{scornet2017tuning}. But, we have seen on our ``toy example'' that estimators proposed herein require few trees in order to have a low estimation error. This makes possible to estimate the indices correctly while maintaining a reasonable computation time.

Besides, we obtain in our application better results for our estimators when comparing with the kernel methods. A major advantage is that we have developed an estimator that requires only one training sample, whereas kernel methods require two training samples or a full one plus a partial. This feature is interesting when dealing with costly models. Another significant asset of our estimators is that their efficiency seems maintained when increasing the model dimension.

Despite these benefits, the proof for the estimators’ consistency as well as the asymptotic analysis to establish the convergence rates and confidence intervals remains a major wish for the future. It is also important to remember that these indices do not have an analogue to the variance decomposition offered by Sobol indices through the theorem of \citet{Hoeffding48}. Thus, using the values of \citet{shapley1953value} could be interesting to get
condensed and easy-to-interpret indices with a good apportionment of the interaction and dependences contributions between the inputs involved.

\clearpage
\bibliographystyle{apalike}
\bibliography{bibliography}

\clearpage
\appendix
\section{Appendix}
\subsection{Algorithms for estimating the first-order QOSA index}
\label{sec:appx:qosa_algorithms}

\begin{algorithm}[h!]
\DontPrintSemicolon
	\KwIn{
		\begin{itemize}
		\item Datasets: $\cD_n^{\diamond} = \left( \bX^{\diamond j}, Y^{\diamond j} \right)_{j= 1,\ldots,n}$ and $\cD_n = \left( \bX^j, Y^j \right)_{j= 1,\ldots,n}$
		\item Number of trees: $k \in \NN^\star$
		\item Order where estimating the QOSA index : $\alpha \in \left] 0, 1 \right[$
		\item Grid where looking for the best parameter: $grid \mathunderscore min \mathunderscore samples \mathunderscore leaf$
		\item Number of folds: $K \in \left\lbrace 2,\ldots,n \right\rbrace$
		\end{itemize}
	}
	\KwOut{Estimated value of the QOSA index at the $\alpha$-order $\widehat{S}_i^\alpha$ for all inputs.}
	
	\vspace{\baselineskip}
	Compute $\widehat{P}$ thanks to Equation \eqref{eq:2:denum_estimator_QOSA_v1}. \;
	\ForEach{$i = 1,\ldots,d$}{
	$\cD_n^{\diamond i} = \left( X_i^{\diamond j}, Y^{\diamond j} \right)_{j= 1,\ldots,n}$ from $\cD_n^{\diamond}$ and $\cD_n^i = \left( X_i^j, Y^j \right)_{j= 1,\ldots,n}$ from $\cD_n$ \;
	Cross-validation as in Algorithm \ref{algo:5:CV_process} with $\cD_n^i$ to get the optimal leaf size $\ell_{opt}$. \;
	Fit a random forest model with $\cD_n^i$ by fixing the $min \mathunderscore samples \mathunderscore leaf$ hyperparameter to $\ell_{opt}$. \;
	Compute the estimator $\widehat{R}_i$ with $\cD_n^{\diamond i}$. \;
	Compute $\widehat{S}_i^\alpha = 1 - \widehat{R}_i / \widehat{P}$. \;
		}
	\caption{QOSA index estimators plugging the quantile}
	\label{algo:appx:QOSA_estimate_two_samples}
\end{algorithm}
\begin{algorithm}[h!]
\DontPrintSemicolon
	\KwIn{
		\begin{itemize}
		\item Datasets: $\cD_n = \left( \bX^{j}, Y^{j} \right)_{j= 1,\ldots,n}$ and $\left( \bX^{\diamond j} \right)_{j= 1,\ldots,n}$
		\item Number of trees: $k \in \NN^\star$
		\item Order where estimating the QOSA index : $\alpha \in \left] 0, 1 \right[$
		\item Grid where looking for the best parameter: $grid \mathunderscore min \mathunderscore samples \mathunderscore leaf$
		\item Number of folds: $K \in \left\lbrace 2,\ldots,n \right\rbrace$
		\end{itemize}
	}
	\KwOut{Estimated value of the QOSA index at the $\alpha$-order $\widehat{S}_i^\alpha$ for all inputs.}
	
	\vspace{\baselineskip}
	Compute $\widehat{P}$ thanks to Equation \eqref{eq:2:denum_estimator_QOSA_v1}. \;
	\ForEach{$i = 1,\ldots,d$}{
	$\cD_n^i = \left( X_i^j, Y^j \right)_{j= 1,\ldots,n}$ from $\cD_n$ and $\left( X_i^{\diamond j} \right)_{j= 1,\ldots,n}$ \;
	Cross-validation as in Algorithm \ref{algo:5:CV_process} with $\cD_n^i$ to get the optimal leaf size $\ell_{opt}$. \;
	Fit a random forest model with $\cD_n^i$ by fixing the $min \mathunderscore samples \mathunderscore leaf$ hyperparameter to $\ell_{opt}$. \;
	Compute the estimator $\widehat{Q}_i \in \left\lbrace \widehat{Q}_i^{1, b}, \widehat{Q}_i^{1, o}\right\rbrace$ with $\cD_n^{\diamond i}$ and $\left( X_i^{\diamond j} \right)_{j= 1,\ldots,n}$. \;
	Compute $\widehat{S}_i^\alpha = 1 - \widehat{Q}_i / \widehat{P}$ \;
		}
	\caption{QOSA index estimators with the weighted minimum approach}
	\label{algo:appx:QOSA_estimate_weighted_min}
\end{algorithm}

\begin{algorithm}[h!]
\DontPrintSemicolon
	\KwIn{
		\begin{itemize}
		\item Datasets: $\cD_n = \left( \bX^{j}, Y^{j} \right)_{j= 1,\ldots,n}$
		\item Number of trees: $k \in \NN^\star$
		\item Order where estimating the QOSA index : $\alpha \in \left] 0, 1 \right[$
		\item Grid where looking for the best parameter: $grid \mathunderscore min \mathunderscore samples \mathunderscore leaf$
		\item Number of folds: $K \in \left\lbrace 2,\ldots,n \right\rbrace$
		\end{itemize}
	}
	\KwOut{Estimated value of the QOSA index at the $\alpha$-order $\widehat{S}_i^\alpha$ for all inputs.}
	
	\vspace{\baselineskip}
	Compute $\widehat{P}$ thanks to Equation \eqref{eq:2:denum_estimator_QOSA_v1}. \;
	\ForEach{$i = 1,\ldots,d$}{
	$\cD_n^i = \left( X_i^j, Y^j \right)_{j= 1,\ldots,n}$ from $\cD_n$ \;
	Cross-validation as in Algorithm \ref{algo:5:CV_process} with $\cD_n^i$ to get the optimal leaf size $\ell_{opt}$. \;
	Fit a random forest model with $\cD_n^i$ by fixing the $min \mathunderscore samples \mathunderscore leaf$ hyperparameter to $\ell_{opt}$. \;
	Compute the estimator $\widehat{Q}_i \in \left\lbrace \widehat{Q}_i^{2, b}, \widehat{Q}_i^{2, o}\right\rbrace$. \;
	Compute $\widehat{S}_i^\alpha = 1 - \widehat{Q}_i / \widehat{P}$ \;
		}
	\caption{QOSA index estimators computing the minimum in leaves}
	\label{algo:appx:QOSA_estimate_min_in_leaf}
\end{algorithm}
\begin{algorithm}[h!]
\DontPrintSemicolon
	\KwIn{
		\begin{itemize}
		\item Datasets: $\cD_n = \left( \bX^{j}, Y^{j} \right)_{j= 1,\ldots,n}$ and $\left( \bX^{\diamond j} \right)_{j= 1,\ldots,n}$
		\item Number of trees: $k \in \NN^\star$
		\item Order where estimating the QOSA index : $\alpha \in \left] 0, 1 \right[$
		\item Minimum number of samples required in a leaf node: $min \mathunderscore samples \mathunderscore leaf \in \{ 1, . . . , n \} $
		\end{itemize}
	}
	\KwOut{Estimated value of the QOSA index at the $\alpha$-order $\widehat{S}_i^\alpha$ for all inputs.}
	
	\vspace{\baselineskip}
	Compute $\widehat{P}$ thanks to Equation \eqref{eq:2:denum_estimator_QOSA_v1}. \;
	Fit a random forest model with $\cD_n$ and the $min \mathunderscore samples \mathunderscore leaf$ hyperparameter. \;
	\ForEach{$i = 1,\ldots,d$}{
	Compute the estimator $\widehat{Q}_i \in \left\lbrace \widehat{Q}_i^{3, b}, \widehat{Q}_i^{3, o}\right\rbrace$ with $\left( \bX^{\diamond j} \right)_{j= 1,\ldots,n}$. \;
	Compute $\widehat{S}_i^\alpha = 1 - \widehat{Q}_i / \widehat{P}$ \;
		}
	\caption{QOSA index estimators with the weighted minimum and fully grown trees}
	\label{algo:appx:QOSA_estimate_weighted_min_and_trees_fully_grown}
\end{algorithm}

\end{document}